\documentclass[a4paper,11pt]{article}
 
 \usepackage{amsmath, amssymb, latexsym, amscd, amsthm,amsfonts,amstext}
 \usepackage[mathscr]{eucal}
 \usepackage{graphicx}
 \usepackage{subfig}
    \newtheoremstyle{example}{\topsep}{\topsep}%
      {}%         Body font
      {}%         Indent amount (empty = no indent, \parindent = para indent)
      {\bfseries}% Thm head font
      {.}%        Punctuation after thm head
      {5pt}%     Space after thm head (\newline = linebreak)
      {\thmname{#1}\thmnumber{ #2}\thmnote{ #3}}%         Thm head spec

    \theoremstyle{example}

\title{Reconstruction from blind  experimental data for an inverse problem for a hyperbolic equation}

\author{Larisa Beilina$^{1\ast}$, Nguyen Trung Th\`anh$^{\circ}$, Michael V. Klibanov$^{\circ}$ 
and Michael A. Fiddy$^{\dagger}$\\
\\
$^{1\ast}$Corresponding author, Department of Mathematical Sciences, \\
Chalmers University of Technology and
Gothenburg University,\\
$^\circ$Department of Mathematics \& Statistics, University of North Carolina at Charlotte,\\
Charlotte 28223, NC, USA (tnguy152@uncc.edu; mklibanv@uncc.edu).\\
 SE-42196 Gothenburg, Sweden (larisa@chalmers.se).\\
$^\dagger$Optoelectronics Center, Univeristy of North Carolina at Charlotte, \\
Charlotte NC 28223, USA. (mafiddy@uncc.edu)}
\date{}

\begin{document}

 \maketitle

\begin{abstract}
We consider the problem of reconstruction of dielectrics from blind
backscattered experimental data.  Experimental data were collected by
a device, which was built at University of North Carolina at
Charlotte. This device sends electrical pulses into the medium and
collects the time resolved backscattered data on a part of a
plane. The spatially distributed dielectric constant $\varepsilon
_{r}\left( \mathbf{x}\right) ,\mathbf{x}\in \mathbb{R}^{3}$ is the
unknown coefficient of a wave-like PDE. This coefficient is
reconstructed from those data in blind cases. To do this, a globally
convergent numerical method is used.\end{abstract}

% -------------------------------------------------------

\textbf{AMS classification codes:} 65N15, 65N30, 35J25.

\graphicspath{{Figures/}}

\section{Introduction}

\label{sec:1}

We consider the problem of reconstruction of refractive indices or
dielectric constants of unknown targets placed in a homogeneous domain
from blind backscattered experimental data. We work with time resolved
backscattering experimental data of wave propagation for a 3-d
hyperbolic coefficient inverse problem (CIP).  Our data are generated
by a single location of the point source. The backscattering signal is
measured on a part of a plane. We present a combination of the
approximately globally convergent method of \cite{BK} with a Finite
Element Method (FEM) for the numerical solution of this CIP. Given a
certain function computed by the technique of \cite{BK}, the FEM
reconstructs the unknown coefficient in an explicit form. As a result,
we can reconstruct refractive indices and locations of targets. In
addition, we estimate their sizes. We believe that these results can
be used as initial guesses for locally convergent methods in order to
obtain better shapes, see, e.g. section 5.9 in \cite{BK}, where the
image obtained by the globally convergent method for transmitted
experimental data was refined via a locally convergent adaptivity
technique.

Experimental data were collected by the device which was recently built at
University of North Carolina at Charlotte. 
%The term \textquotedblleft blind
%data" means here that the mathematical team (NTT,LB,MVK) did not know
%correct answers in advance. This team has presented computational results to
%MAF, who oversaw the data collection process.\ Next, MAF has compared
%computational results with directly measured ones (section 8).\ Hence, the
%blind data case is unbiased. 
In our experiments we image targets standing in the air. A potential
application of our experiments is in imaging of explosives. Note that
explosives can be located in the air \cite{KBKSNF}, e.g. improvised
explosive devices (IEDs). The work on real data for the case when
targets are hidden in a soil is ongoing.

We have collected backscattering time resolved experimental data of
electrical waves propagation in a non-attenuating medium. As it was pointed
out in \cite{BK,KBKSNF}, the main difficulty of working with such data is
caused by a huge mismatch between these data and ones produced by
computational simulations. Conventional data denoising techniques do not
help in this case. Therefore, it is unlikely that any numerical method would
successfully invert the raw data. To get the data, which would look somewhat
similar with ones obtained in computational simulations, a heuristic data
pre-processing procedure should be applied. The pre-processed data are used
as the input for the globally convergent method.

Previously our research group has applied the method of \cite{BK} to
the simpler case of transmitted experimental data which were produced
by a similar device (chapter 5 of \cite{BK}). The backscattering real
data are much harder to work with than transmitted ones since the
backscattered signal is significantly weaker than the transmitted one,
as well as because some unwanted signals are mixed up with the true
one, see Figure \ref{fig2}-a) for the latter.  We refer to our
research in \cite{KBKSNF} and section 6.9 of \cite{BK} for the case of
backscattering real data in 1-d.  In the current paper we present
results of reconstruction of the 3-d version of the method of \cite{BK}.

The approximately globally convergent method of \cite{BK} relies on
the structure of the underlying PDE operator and does not use
optimization techniques. Each iterative step consists of solutions of
two problems: the Dirichlet boundary value problem for an elliptic PDE
and the Cauchy problem for the underlying hyperbolic
PDE. \textquotedblleft Approximate global convergence" (global
convergence in short) means that we use a certain reasonable
approximate mathematical model. Approximation is used because of one
inevitably faces with substantial challenges when trying to develop
globally convergent numerical methods for multidimensional CIPs for
hyperbolic PDEs with single source. It is rigorously established in
the framework of this model that the method of \cite{BK} results in
obtaining some points in a small neighborhood of the exact coefficient
without a priori knowledge of any point in this neighborhood, see
Theorem 2.9.4 in \cite{BK} and Theorem 5.1 in \ \cite{BKJIIP12}. The
distance between those points and the exact solution depends on the
error in the data, the step size $h$ of a certain discretization of
the pseudo-frequency interval and the computational domain $\Omega $
where the inverse problem is solved (see section \ref{sec:4.3} for
definition of $h$). A knowledge of the background medium in $\Omega$
is also not required by this method. Because of these theorems,
convergence analysis is not presented here. A rigorous definition of
the approximate global convergence property can be found in section
1.1.2 of \cite{BK} and in \cite{BKJIIP12}. We use a mild
approximation, since it amounts only to the truncation of a certain
asymptotic series, and it is used only on the first iterative step
(section \ref{sec:4.2}). The validity of this approximate model was
verified computationally on both synthetic and transmitted
experimental data in \cite{BK,BKJIIP12} as well as in the current work
in the case of experimental backscattering data.

Different imaging methods are used to compute geometrical information
of targets, such as their shapes, sizes and locations, see,
e.g.~\cite{Isakov,Soumekh}. On the other hand, refractive indices,
which is our main interest, characterize constituent materials of
targets, and they are much more difficult to compute. As to the
gradient-like methods, we refer to, e.g. \cite{BakKok,Chav,EHN,Vog}
and references therein. Convergence of these methods is guaranteed
only if the starting point of iterations is chosen to be sufficiently
close to the correct solution. On the other hand, it was shown in
section 5.8.4 of \cite{BK} that the gradient method failed to work for
transmitted experimental data of \cite{BK} in the case when its
starting point was the background medium.

An outline of this paper is as follows.  In section 2 we state forward
and inverse problems. In section 3 we describe the experimental data
and briefly outline the data pre-processing procedure.  In section 4
we briefly outline the method of \cite{BK}: for reader's
convenience. In section 5 we describe a version of the FEM which works
for our case. In section 6 we describe our algorithm. In section 7 we
outline some details of our numerical implementation.\ Results are
presented in section 8 and summary is in section 9.

\section{Statements of Forward and Inverse Problems}

\label{sec:2}

Let $\Omega \subset \mathbb{R}^{3}$ be a convex bounded domain with the
boundary $\partial \Omega \in C^{3}.$ Denote by $\mathbf{x}=\left( x,y,z\right)
\in \mathbb{R}^{3}.$ We model the electromagnetic wave propagation in an
isotropic and non-magnetic space $\mathbb{R}^{3}$ with the dimensionless
coefficient $\varepsilon _{r}(\mathbf{x}),$ which describes the spatially
distributed dielectric constant of the medium. We consider the following
Cauchy problem for the hyperbolic equation 
\begin{equation}
\varepsilon _{r}(\mathbf{x})u_{tt}=\Delta u\text{ in }\mathbb{R}^{3}\times
\left( 0,\infty \right) ,  \label{2.1}
\end{equation}%
\begin{equation}
u\left( \mathbf{x},0\right) =0,\text{ }u_{t}\left( \mathbf{x},0\right)
=\delta \left( \mathbf{x}-\mathbf{x}_{0}\right) .  \label{2.2}
\end{equation}%
We assume that the coefficient $\varepsilon _{r}(\mathbf{x})$ of equation (%
\ref{2.1}) is such that 
\begin{equation}
\varepsilon _{r}(\mathbf{x})\in C^{\alpha }\left( \mathbb{R}^{3}\right)
,\varepsilon _{r}(\mathbf{x})\in \lbrack 1,b],~~\varepsilon _{r}(\mathbf{x}%
)=1\text{ for }\mathbf{x}\in \mathbb{R}^{3}\diagdown \Omega ,  \label{2.20}
\end{equation}%
where $b=const.>1.$ We \emph{a priori} assume knowledge of the constant $b,$
which amounts to the knowledge of the set of admissible coefficients in (\ref%
{2.20}). However, we do not assume that the number $b-1$ is small, i.e. we
do not impose smallness assumptions on the unknown coefficient $\varepsilon
_{r}(\mathbf{x})$. Below $C^{k+\alpha }$ are H\"older spaces, where $k\geq 0$
is an integer and $\alpha \in \left( 0,1\right) .$ Let $\Gamma \subset
\partial \Omega $ be a part of the boundary $\partial \Omega .$ Later we
will designate $\Gamma $ as the backscattering side of $\Omega $ and will
explain how we deal with the absence of the data at $\partial \Omega
\setminus \Gamma .$

\textbf{Coefficient Inverse Problem (CIP).} \emph{Suppose that the
coefficient }$\varepsilon _{r}\left( \mathbf{x}\right) $\emph{\ satisfies %
\eqref{2.20}. Determine the function }$\varepsilon _{r}\left( \mathbf{x}%
\right) $\emph{\ for }$\mathbf{x}\in \Omega $\emph{, assuming that the
following function }$g(\mathbf{x},t)$\emph{\ is known for a single source
position }$x_{0}\notin \overline{\Omega }$ 
\begin{equation}
u\left( \mathbf{x},t\right) =g\left( \mathbf{x},t\right) ,\forall \left( 
\mathbf{x},t\right) \in \Gamma \times \left( 0,\infty \right) .  \label{2.5}
\end{equation}

The function $g(\mathbf{x},t)$ in (\ref{2.5}) models time dependent
measurements of the wave field at the part $\Gamma $ of the boundary $%
\partial \Omega $ of the domain of interest $\Omega $. We assume below that
the source position is fixed and $\mathbf{x}_{0}\notin \overline{\Omega }$.
This assumption allows us to simplify the resulting integral-differential
equation because $\delta (\mathbf{x}-\mathbf{x}_{0})=0$ in $\overline{\Omega 
}$. The assumption $\varepsilon _{r}(\mathbf{x})=1$ for $x\in \mathbb{R}%
^{3}\diagdown \Omega $ means that the coefficient $\varepsilon _{r}(\mathbf{x%
})$ has a known constant value outside of the domain of interest $\Omega .$

This is a CIP with single measurement data. Uniqueness theorem for such CIPs
in the multidimensional case are currently known only if the function $%
\delta \left( \mathbf{x}-\mathbf{x}_{0}\right) $ in (\ref{2.2}) is replaced
with a function $f\left( \mathbf{x}\right) $ such that $\Delta f\left( 
\mathbf{x}\right) \neq 0$ $\forall \mathbf{x}\in \overline{\Omega }.$ A
proper example of such function $f$ is a narrow Gaussian centered around $%
\mathbf{x}_{0}$, which approximates the function $\delta \left( \mathbf{x}-%
\mathbf{x}_{0}\right) $ in the distribution sense.\ From the Physics
standpoint this Gaussian is equivalent to $\delta \left( \mathbf{x}-\mathbf{x%
}_{0}\right) .$ That uniqueness theorem can be proved by the method, which
was originated in \cite{BukhK}. This method is based on Carleman estimates,
also see, e.g. sections 1.10, 1.11 of the book \cite{BK} about this method.
The authors believe that, because of applications, it still makes sense to
develop numerical methods for this CIP without completely addressing the
uniqueness question.

The function $u\left( \mathbf{x},t\right) $ in (\ref{2.1}) represents the
voltage of one component of the electric field $E\left( \mathbf{x},t\right)
=\left( E_{x},E_{y},E_{z}\right) \left( \mathbf{x},t\right).$ In our
  computer simulations the incident field has only one non-zero component $E_{z}$. This
component propagates along the $z-$axis until it reaches the target, where
it is scattered. So, we assume that in our experiment  $u\left( \mathbf{x},t\right) =E_{z}\left( \mathbf{x}
,t\right) .$
% "Electric wave" is used here to mean the component $E_{z}\left( 
%\mathbf{x},t\right) $ of the electromagnetic wave generated by our microwave
%source.
 We now comment on five main discrepancies between our mathematical
model (\ref{2.1})- (\ref{2.20}) and the reality. The first discrepancy which
causes the main difficulties, is the aforementioned huge mismatch between
experimental data and computational simulations. The second one is that,
although we realize that equation (\ref{2.1}) can be derived from Maxwell
equations only in the 2-d case, we use it to model the full 3-d case. The
reason is that our current receiver can measure only one of the polarization
components of the scattered electric field $E$. In addition, if using a more
complicated mathematical model than the one of (\ref{2.1}), for example the
one that includes vector scattering and thus depolarization effects on
scattering, then one would need to develop a globally convergent inverse
method for this case. The latter is a quite time consuming task with yet
unknown outcome.\textbf{\ }Equation (\ref{2.1}) was used in Chapter 5 of 
\cite{BK} for the case of transmitted experimental data, and accurate
solutions were obtained. A partial explanation of the latter can be found in 
\cite{BM}, where the Maxwell's system in a non-magnetic and non-conductive
medium was solved numerically in time domain.\ It was shown numerically in
section 7.2.2 of \cite{BM} that the component of the vector $E\left( \mathbf{%
x},t\right) =\left( E_{x},E_{y},E_{z}\right) \left( \mathbf{x},t\right) ,$
which was initially incident upon the medium, dominates two other
components. This is true for at least a rather simple medium such as ours.
Therefore, the function $u\left( \mathbf{x},t\right) $ in (\ref{2.1})
represents the voltage of the computed component $E_{z}\left( \mathbf{x},t\right)$ 
of the electric field, which is emitted and measured by our antennas.

The third discrepancy is that the condition $\varepsilon _{r}(\mathbf{x})\in
C^{3}\left( \mathbb{R}^{3}\right) $ is violated on the inclusion/background
interface in our experiments. The fourth discrepancy is that formally
equation (\ref{2.1}) is invalid for the case when metallic targets are
present. On the other hand, it was demonstrated computationally in \cite%
{KBKSNF} that one can treat metallic targets as dielectrics with large
dielectric constants, which we call \emph{appearing dielectric constant}, 
\begin{equation}
\varepsilon _{r}\left( \text{metallic target}\right) \in \left( 10,30\right)
.  \label{2.51}
\end{equation}%
Modeling metallic targets as integral parts of the unknown coefficient $%
\varepsilon _{r}\left( \mathbf{x}\right) $ is convenient for the above
application to imaging of explosives. Indeed, IEDs usually consist of
mixtures of some dielectrics with a number of metallic parts. Such targets
are heterogeneous ones, and we consider three heterogeneous cases in section %
\ref{sec:8.2}. On the other hand, modeling metallic parts of heterogeneous
targets as a separate matter than the rest of an a priori unknown background
medium would result in significant additional complications of the already
difficult problem with yet unknown outcome.

The fifth discrepancy is that we use the incident plane wave instead of the
point source in our computations. We have discovered that the plane wave
case works better in image reconstructions than the point source, while the
point source case is more convenient for the convergence analysis in \cite%
{BK,BKJIIP12}. In addition, since the distance between our measurement plane
and targets is much larger than the wavelength of our signal, it is
reasonable to approximate the incident wave as a plane wave.

Thus, our results of section \ref{sec:8.2} demonstrate the well known fact
that computational results are often less pessimistic than the theory, since
the theory cannot grasp all nuances of the reality. In summary, we believe
that accurate solutions of the above CIP for experimental data justify our
mathematical model.

\section{Experimental Data}

% 
% \begin{figure}[h]
% \begin{center}
% \begin{tabular}{cc}
% \multicolumn{2}{c}
% {a) \includegraphics[scale=0.2]{F1a.ps}} \\
%   {\includegraphics[scale=0.4]{F1b.ps}} & 
%  {\includegraphics[scale=0.22]{F1c.ps}} \\
% b)  & c)
% \end{tabular}
% \end{center}
% \caption{\emph{a) A photograph explaining our data collection process. The
% distance between the target (wooden block) and the measurement plane is
% about 0.8 m, which is about 26 wave lengths. b) Picosecond Pulse Generator. c) Textronix Oscilloscope.  }}
% \label{fig1}
% \end{figure}

\begin{figure}[h]
\begin{center}
 \subfloat[]{\includegraphics[scale=0.45]{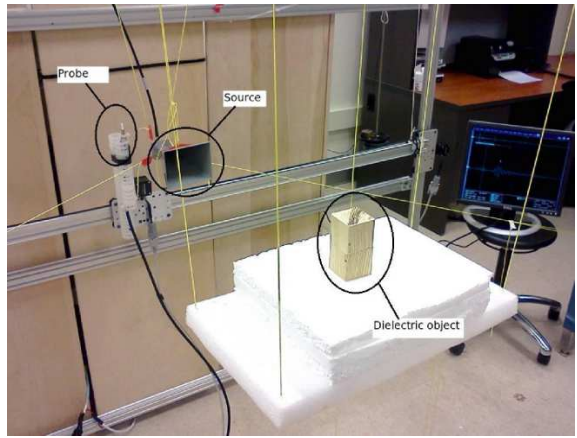}} 

 \subfloat[]{\includegraphics[scale=0.4]{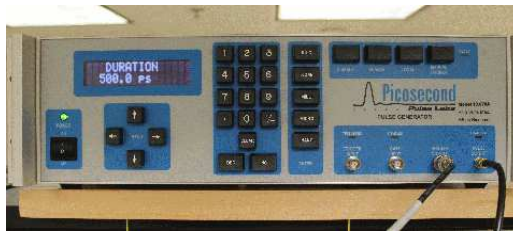}}  
 \subfloat[]{\includegraphics[scale=0.3]{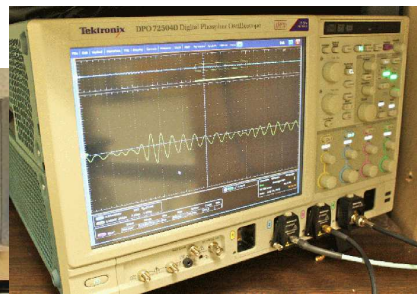}}
\end{center}
\caption{\emph{a) A photograph explaining our data collection process. The
distance between the target (wooden block) and the measurement plane is
about 0.8 m, which is about 26 wave lengths. b) Picosecond Pulse Generator. c) Textronix Oscilloscope.  }}
\label{fig1}
\end{figure}

\subsection{Data collection}

\label{sec:3.1}

Figure \ref{fig1}-a) is a photograph explaining the data collection. The data
collection is done in a regular room, which contains office furniture,
computers, etc. Keeping in mind our desired application (see Introduction),
we intentionally did not arrange a special waveguide, which would protect
our data from unwanted signals caused by reflections from various objects in
the room. Below $x$ and $y$ are horizontal and vertical axis respectively
and the $z$ axis is perpendicular to the measurement plane, the positive
direction of $z$ axis is in the direction from the target to the measurement
plane. We dimensionalize our coordinates as $\mathbf{x}^{\prime }=\mathbf{x}/(1m),$ where \textquotedblleft m" stands for meter. However, we do not
change notations of coordinates for brevity. Hence, below, e.g. 0.05 of
length actually means 5 centimeters.

%\begin{figure}[p]
%\centering
%\subfloat[]{\includegraphics[width = 0.45\textwidth]{figure/F3a}} %
%\subfloat[]{\includegraphics[width = 0.45\textwidth]{figure/F3b}}
%\caption{ \emph{a) A typical 2-d distribution of the measured function }$%
%u_{meas}\left( \mathbf{x},t\right) $\emph{\ at a certain moment of time }$%
%t=t_{0}$\emph{\ at the measurement plane. b) The 2-d distribution of the
%computationally simulated function }$u_{comp}\left( \mathbf{x},t\right) $%
%\emph{\ for the same target at }$t=t_{0}$\emph{\ at the measurement plane.
%Again, a significant difference between a) and b) can be observed.}}
%\label{fig3}
%\end{figure}

"The transmitter sends the pulse into the medium which contains targets of interest.  The electric wave caused by the
pulse is scattered by the targets, and the backscattered signal is detected by the detector. The detected signal is recorded
by the real time oscilloscope."

Two main pieces of our device are Picosecond Pulse Generator (Figure
\ref{fig1}-b)) and Textronix Oscilloscope (Figure \ref{fig1}-c)). The
Picosecond Pulse Generator generates electric pulses. The duration of
each pulse is 300 picoseconds. This pulse goes to the transmitter,
which is a horn antenna (source). 

The transmitter sends the pulse into the medium which contains targets
of interest.  The electric wave caused by the pulse is scattered by
the targets, and the backscattered signal is detected by the
detector. The detected signal is recorded by the real time
oscilloscope. The oscilloscope produces a digitized time resolved
signal with the step size in time of 10 picoseconds. The total time of
measurements for one pulse is 10 nanoseconds=10$^{4}$
picoseconds=10$^{-8}$ second.

To decrease the measurement noise, the pulse is generated 800 times for each
position of the detector, the backscattering signal is also measured 800
times and resulting signals are averaged. The detector moves in both
horizontal and vertical directions covering the square $SQ=\left\{
-0.5<x,y<0.5\right\} $ on the measurement plane. We have chosen the step
size of this movement to be 0.02. Although we can choose any step size, we
found that 0.02 provides a good compromise between the precision of
measurements and the total time spent on data collection.

The distance between our targets and the measurement plane is approximately
0.8 with about 0.05 deviations, and the wavelength of our signal is about
0.03. Therefore, the distance between the measurement plane and our targets
is of about 26 wavelengths. This is in the far field zone.

\subsection{Data pre-processing}

\label{sec:3.2}

\begin{figure}[h]
\begin{center}
 \subfloat[]{\includegraphics[width = 0.5\textwidth]{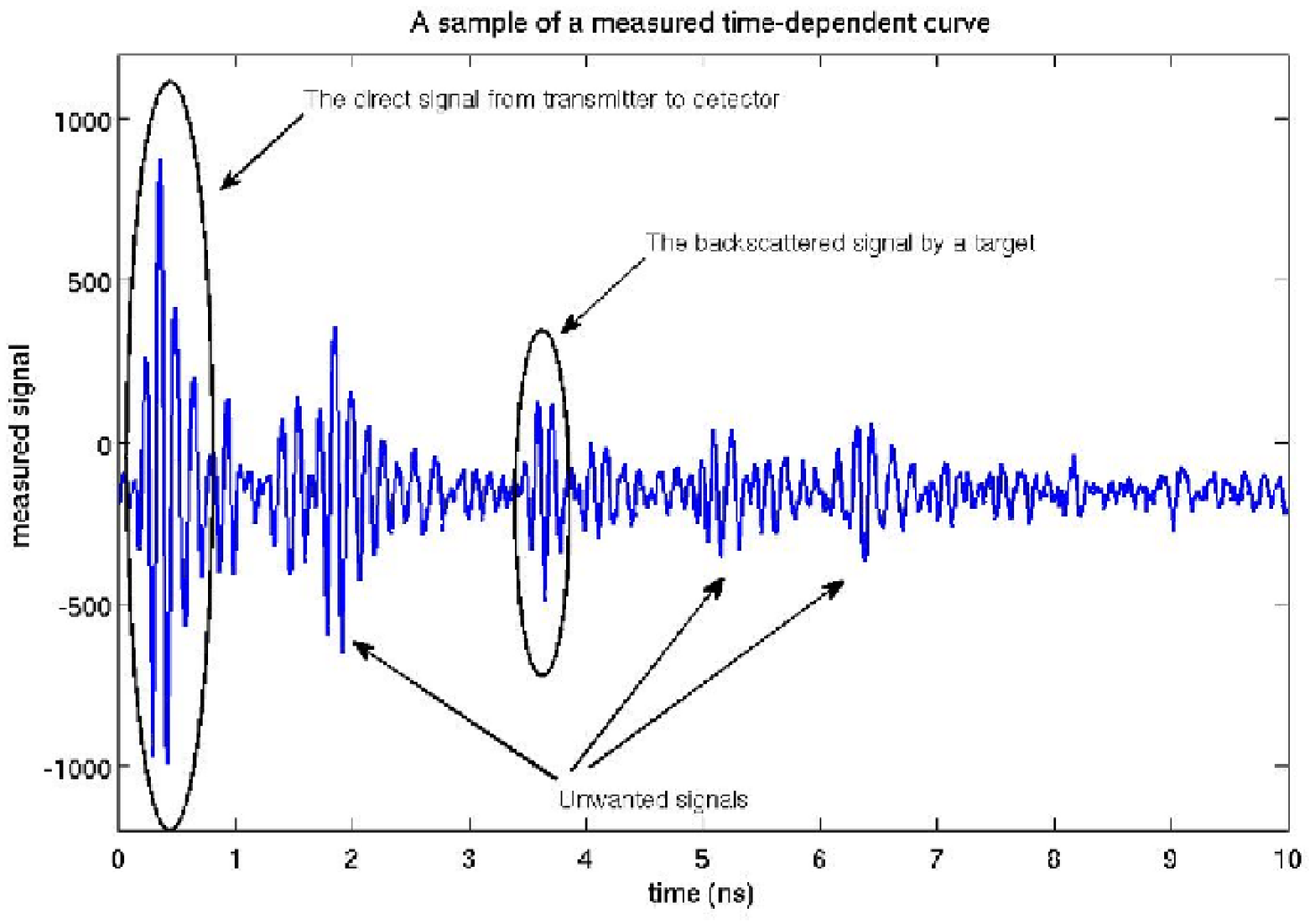}} 
 \subfloat[]{\includegraphics[width = 0.5\textwidth]{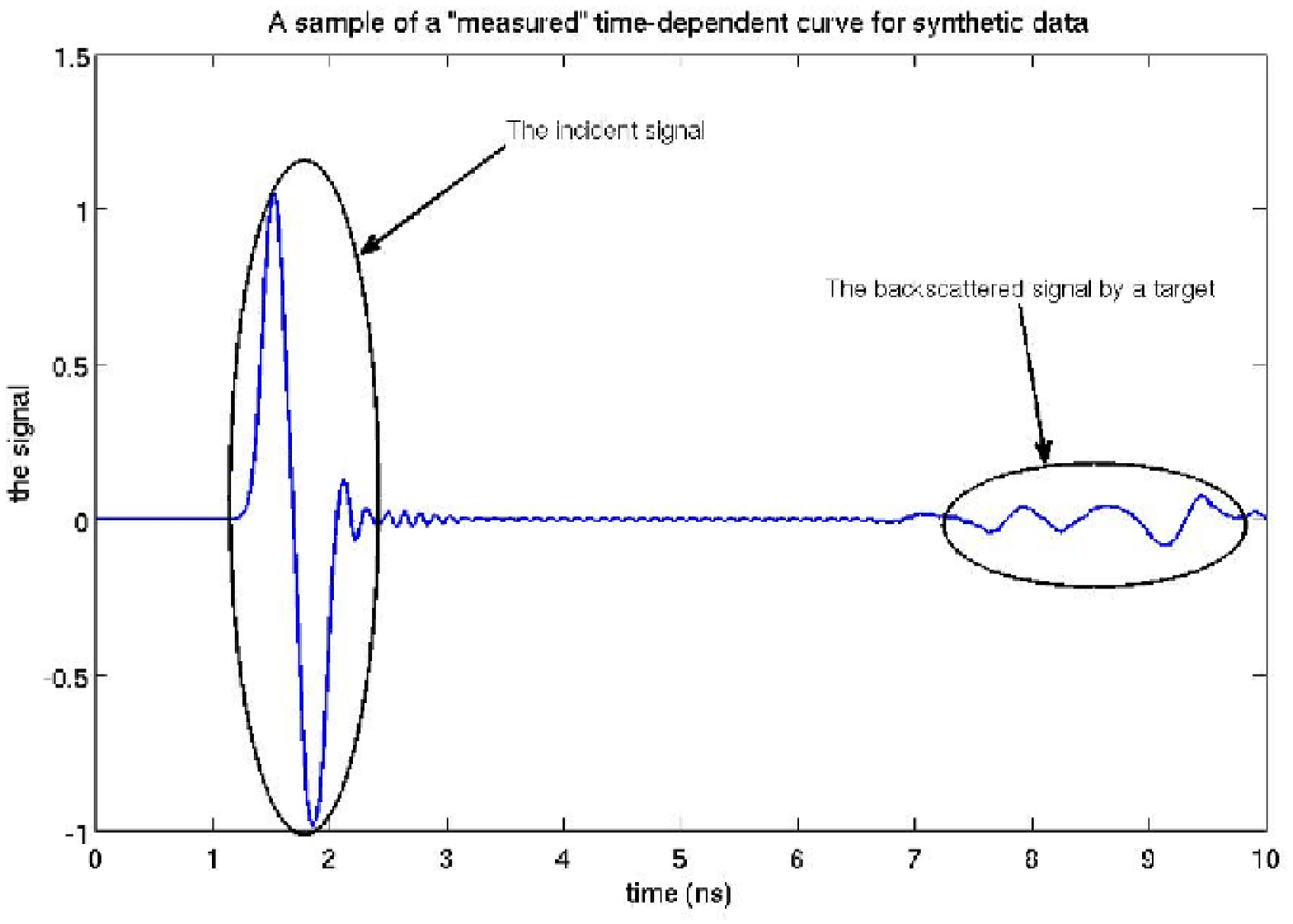}}
\end{center}
\caption{\emph{Typical samples of real and computationally simulated data.
a) The measured data at one of detectors. The direct signal goes from the
transmitter (Fig.~\protect\ref{fig1}-b)) to the detector because the
transmitter emits the electric field in all directions. We use the direct
signal as the time reference in our data pre-processing procedure. Unwanted
signals are due to reflections from a variety of objects in the room. b) The
computationally simulated data for the same target as the one of a) and at
the same detector. A significant difference between a) and b) is evident.}}
\label{fig2}
\end{figure}

The main difficulty working with experimental data is that there is a
huge mismatch between these data and computationally simulated
ones. Indeed, Figure \ref{fig2}-a) depicts a sample of experimentally
measured data for a wooden block at one position of the detector, see
Figure \ref{fig1}-a) for data collection scheme. On this figure, the
direct signal is the signal going directly to the receiver.
 We use this direct signal as the time reference for data
pre-processing. Unwanted signals are due to reflections of the electric wave
from several objects present in the room. Figure \ref{fig2}-b) presents the
computationally simulated data for the same target, see section \ref{sec:7.1}
for data simulations. These figures show a huge mismatch between real and
computationally simulated data. 
%In addition, this mismatch is evident from
%comparison of Figures \ref{fig3}-a) and \ref{fig3}-b) which show the
%distributions on the measurement plane of the experimental and simulated
%data for the same target at a certain moment of time $t=t_{0}.$
Therefore, data pre-processing is necessary. We refer to \cite{NBKF} for
details of our data pre-processing procedure. The main steps of this
procedure include:

\begin{enumerate}
\item \emph{Time-zero correction}. The time-zero correction is to shift the
measured data in time. So that its starting time is the same as when the
incident pulse is emitted from the transmitter. This is done using the
direct signals from the transmitter to the detector as the time reference.

\item \emph{Extraction of scattered signals}. Apart from the backscattered
wave by the targets, our measured data also contain various types of
signals, e.g.~direct signals from the horn to the detector, scattered
signals from structures inside the room, etc. What we need, however, is the
scattered signals by the targets only. To obtain them, we single out the
scattered signals caused by the targets only and remove all unwanted signals.

\item \emph{Data propagation}. After getting the scattered signals, the next
step of data pre-processing is to propagate the data closer to the targets,
i.e.~to approximate the scattered wave on a plane which is much closer to
the targets then the measurement plane. The distance between that propagated
plane and the front surface of a target is usually between 0.02 and 0.06
(compare with the 0.8 distance from the measurement plane). There are two
reasons for doing this. The first one is that the method of \cite{BK} works
with the Laplace transform of the function $u\left( \mathbf{x},t\right) $
(section \ref{sec:4}). That Laplace transform decays exponentially in terms
of the time delay, which is proportional to the distance from the target to
the measurement plane.\ Hence, the amplitude of the Laplace transformed
experimental data on the measurement plane is very small and can be
dominated by computational round-off error. The second reason is that this
propagation procedure helps to substantially reduce the computational cost
since the computational domain for the inverse problem is reduced.

\item \emph{Data calibration.} Finally, since the amplitudes of the
experimental incident and scattered waves are usually significantly
different from simulations, we need to bring the former to the same level of
the amplitude as the latter. This is done using a known target referred to
as \emph{calibrating object}.
\end{enumerate}

In this paper, the result of data pre-processing is used as the
measured data $g\left( \mathbf{x},t\right) $ on the backscattering
boundary $\Gamma $ of our computational domain $\Omega $ for the
inverse problem.
% Figure \ref{fig4} compares the Laplace transform of
%the propagated and simulated data on the propagated plane after the
%calibration. The similarity in the behavior of their negative peaks,
%which are due to the target, as well as comparison with the Laplace
%transforms of real and simulated data on the measurement plane confirm
%the usefulness of the data pre-processing.

\section{The Approximately Globally Convergent Method in Brief}

\label{sec:4}

In this section we briefly outline the globally convergent method for
reader's convenience. We refer to sections 2.3, 2.5, 2.6.1 and 2.9.2 of \cite%
{BK} as well as \cite{BKJIIP12} for details.

The first step of our inverse algorithm is the Laplace transform of the
function $u\left( \mathbf{x},t\right) ,$%
\begin{equation}
w(\mathbf{x},s)=\int\limits_{0}^{\infty }u(\mathbf{x},t)e^{-st}dt,\text{ for 
}s>\underline{s}=const.>0,  \label{2.6}
\end{equation}%
where $\underline{s}$ is a certain number. We assume that the number $%
\underline{s}$ is sufficiently large, and we call the parameter $s$ \emph{%
pseudo frequency}. It follows from (\ref{2.1}), (\ref{2.2}) and (\ref{2.6})
that the function $w$ is the solution of the following problem 
\begin{equation}
\Delta w-s^{2}\varepsilon _{r}(\mathbf{x})w=-\delta \left( \mathbf{x}-%
\mathbf{x}_{0}\right) ,\text{ }x\in \mathbb{R}^{3},  \label{2.7}
\end{equation}%
\begin{equation}
\lim_{\left\vert x\right\vert \rightarrow \infty }w\left( \mathbf{x}%
,s\right) =0.  \label{2.8}
\end{equation}%
The limit (\ref{2.8}) is proved in Theorem 2.7.1 of \cite{BK}. In addition,
it was proven in Theorem 2.7.2 of \cite{BK} that for the function $%
\varepsilon _{r}\left( \mathbf{x}\right) $ satisfying (\ref{2.20}) there
exists unique solution $w\left( \mathbf{x},s\right) $ of the problem (\ref%
{2.7}), (\ref{2.8}) for every $s>0$ such that 
\begin{equation*}
w\left( \mathbf{x},s\right) =w_{0}\left( \mathbf{x},s\right) +\overline{w}%
\left( \mathbf{x},s\right) ,\overline{w}\left( \mathbf{x},s\right) \in
C^{2+\alpha }\left( \mathbb{R}^{3}\right) ,
\end{equation*}%
where $w_{0}\left( \mathbf{x},s\right) $ is the solution of the problem (\ref%
{2.7}), (\ref{2.8}) for the case $\varepsilon _{r}(\mathbf{x})\equiv 1,$%
\begin{equation*}
w_{0}\left( \mathbf{x},s\right) =\frac{\exp \left( -s\left\vert \mathbf{x}-%
\mathbf{x}_{0}\right\vert \right) }{4\pi \left\vert \mathbf{x}-\mathbf{x}%
_{0}\right\vert }.
\end{equation*}

\subsection{The integral differential equation}

\label{sec:4.1}

It follows from Theorem 2.7.2 of \cite{BK} that $w(\mathbf{x},s)>0.$ Hence,
we can consider the functions $v(\mathbf{x},s)$, $q(\mathbf{x},s),$ 
\begin{equation}
v\left( \mathbf{x},s\right) =\frac{\ln w\left( \mathbf{x},s\right) }{s^{2}}%
,\ q\left( \mathbf{x},s\right) =\frac{\partial v\left( \mathbf{x},s\right) }{%
\partial s}.  \label{transform}
\end{equation}%
Substituting $w=\exp \left( s^{2}v\right) $ in (\ref{2.7}) and keeping in
mind that the source $\mathbf{x}_{0}\notin \overline{\Omega }$, we obtain 
\begin{equation}
\Delta v+s^{2}|\nabla v|^{2}=\varepsilon _{r}(\mathbf{x}),\mathbf{x}\in
\Omega .  \label{new3.2}
\end{equation}%
Using (\ref{transform}) we obtain 
\begin{equation}
v\left( \mathbf{x},s\right) =-\int\limits_{s}^{\overline{s}}q\left( \mathbf{x%
},\tau \right) d\tau +V\left( \mathbf{x},\overline{s}\right) ,  \label{3.7}
\end{equation}%
where the truncation pseudo frequency $\overline{s}>\underline{s}$ is a
large number, which is chosen numerically, see section \ref{sec:8} for
details. We call $V\left( \mathbf{x},\overline{s}\right) $ the \emph{tail
function,} and it is unknown. It follows from (\ref{transform}) and (\ref%
{3.7}) that 
\begin{equation}
V\left( \mathbf{x},\overline{s}\right) =v\left( \mathbf{x},\overline{s}%
\right) =\frac{\ln w\left( \mathbf{x},\overline{s}\right) }{\overline{s}^{2}}%
.  \label{2.A}
\end{equation}%
It follows from \cite{BK} (section 2.3) that, under some conditions, there
exists a function $p\left( \mathbf{x}\right) \in C^{2+\alpha }\left( 
\overline{\Omega }\right) $ such that the following asymptotic behavior with
respect to $\overline{s}\rightarrow \infty $ holds for functions $V$ and $q$ 
\begin{equation}
V\left( \mathbf{x},\overline{s}\right) =\frac{p\left( \mathbf{x}\right) }{%
\overline{s}}+O\left( \frac{1}{\overline{s}^{2}}\right) ,\text{ }\overline{s}%
\rightarrow \infty ,  \label{2.B}
\end{equation}%
\begin{equation}
q\left( \mathbf{x},\overline{s}\right) =\partial _{\overline{s}}V\left( 
\mathbf{x},\overline{s}\right) =-\frac{p\left( \mathbf{x}\right) }{\overline{%
s}^{2}}+O\left( \frac{1}{\overline{s}^{3}}\right) ,\text{ }\overline{s}%
\rightarrow \infty .  \label{2C}
\end{equation}
Differentiating both sides of equation (\ref{new3.2}) with respect to $s$ then using (\ref{transform}) and (\ref{3.7}), we obtain the following
nonlinear integral differential equation
\begin{equation}
\begin{split}
& \Delta q-2s^{2}\nabla q\int\limits_{s}^{\overline{s}}\nabla q\left( 
\mathbf{x},\tau \right) d\tau +2s\left( \int\limits_{s}^{\overline{s}}\nabla
q\left( \mathbf{x},\tau \right) d\tau \right) ^{2} \\
& +2s^{2}\nabla q\nabla V-4s\nabla V\int\limits_{s}^{\overline{s}}\nabla
q\left( \mathbf{x},\tau \right) d\tau +2s\left( \nabla V\right) ^{2}=0,%
\mathbf{x}\in \Omega ,s\in \left[ \underline{s},\overline{s}\right] .
\end{split}
\label{3.8}
\end{equation}%
In addition, (\ref{2.5}) and (\ref{transform}) lead to the following
Dirichlet boundary condition for the function $q$  
\begin{equation}
q\left( \mathbf{x},s\right) =\widetilde{\psi }\left( \mathbf{x},s\right) ,%
\text{ }\forall \left( \mathbf{x},s\right) \in \Gamma \times \left[ 
\underline{s},\overline{s}\right] ,  \label{3.9}
\end{equation}%
\begin{equation}
\widetilde{\psi }\left( \mathbf{x},s\right) =\frac{\partial _{s}\left( \ln
\varphi \right) }{s^{2}}-2\frac{\ln \varphi }{s^{3}}.  \label{3.90}
\end{equation}%
Here $\varphi \left( \mathbf{x},s\right) $ is the Laplace transform (\ref%
{2.6}) of the function $g\left( \mathbf{x},t\right) $ in (\ref{2.5}). We now
need to complement the boundary data (\ref{3.9}) at the backscattering side $%
\Gamma $ with the boundary data at the rest of the boundary $\partial \Omega
.$ Using computationally simulated data, it was shown numerically in section
6.8.5 of \cite{BK} as well as in \cite{BKJIIP12} that it is reasonable to
approximate the boundary data on $\partial \Omega \setminus \Gamma $ by the
solution of the forward problem for the homogeneous medium for the case $%
\varepsilon _{r}\left( \mathbf{x}\right) =1$: recall that this equality
holds outside of the domain $\Omega ,$ see (\ref{2.20}). Thus, we use below
the following Dirichlet boundary condition for the function $q\left( \mathbf{%
x},s\right) $%
\begin{equation}
q\left( \mathbf{x},s\right) =\psi \left( \mathbf{x},s\right) ,\text{ }%
\forall \left( \mathbf{x},s\right) \in \partial \Omega \times \left[ 
\underline{s},\overline{s}\right] ,  \label{3.91}
\end{equation}%
\begin{equation}
\psi \left( \mathbf{x},s\right) =\left\{ 
\begin{array}{c}
\widetilde{\psi }\left( \mathbf{x},s\right) ,\text{ }\forall \left( \mathbf{x%
},s\right) \in \Gamma \times \left[ \underline{s},\overline{s}\right] , \\ 
\psi ^{0}\left( \mathbf{x},s\right) ,\forall \left( \mathbf{x},s\right) \in
\left( \partial \Omega \setminus \Gamma \right) \times \left[ \underline{s},%
\overline{s}\right] .%
\end{array}%
\right.  \label{3.92}
\end{equation}%
where the function $\psi ^{0}\left( \mathbf{x},s\right) $ is the function $%
\widetilde{\psi }\left( \mathbf{x},s\right) $ in (\ref{3.90}) computed for
the case $\varepsilon _{r}(\mathbf{x})\equiv 1$.

Even though equation (\ref{3.8}) with the boundary condition (\ref{3.91})
has two unknown functions $q$ and $V$, we can approximate both of them
because approximation procedures for them are different, see section 7.1.
Suppose for a moment that functions $q$ and $V$ are approximated in $\Omega $
together with their derivatives $D_{\mathbf{x}}^{\alpha }q,D_{\mathbf{x}%
}^{\alpha }V,\left\vert \alpha \right\vert \leq 2.$ Then the corresponding
approximation for the coefficient $\varepsilon _{r}(\mathbf{x})$ can be
found via backwards calculation using (\ref{new3.2}).

\subsection{The first approximation for the tail function}

\label{sec:4.2}

To start iterations, we need the first approximation $V_{1,0}\left( \mathbf{x%
}\right) $ for the tail function. In this section we show how to calculate $%
V_{1,0}\left( \mathbf{x}\right) .$ This is the same choice as the one in
section 2.9.2 of the book \cite{BK} as well as in \cite{BKJIIP12}.

Let the function $\varepsilon _{r}^{\ast }(\mathbf{x})$ satisfying (\ref%
{2.20}) be the exact solution of our CIP for the exact data $g^{\ast }$ in (%
\ref{2.5}). Let $V^{\ast }\left( \mathbf{x},\overline{s}\right) $ be the
exact \textquotedblleft tail function\textquotedblright\ defined as 
\begin{equation}
V^{\ast }\left( \mathbf{x},\overline{s}\right) =\frac{\ln w^{\ast }\left( 
\mathbf{x},\overline{s}\right) }{\overline{s}^{2}}.  \label{4.17*}
\end{equation}%
Let $q^{\ast }\left( \mathbf{x},s\right) \in C^{2+\alpha }\left( \overline{%
\Omega }\right) \times C\left[ \underline{s},\overline{s}\right] $ be the
corresponding exact function $q\left( \mathbf{x},s\right) $ satisfying
equation (\ref{3.8}). Let $\psi ^{\ast }\left( \mathbf{x},s\right) \in
C^{2+\alpha }\left( \overline{\Omega }\right) \times C\left[ \underline{s},%
\overline{s}\right] $ be the corresponding exact Dirichlet boundary
condition for $q^{\ast }\left( \mathbf{x},s\right) $ as defined in (\ref%
{3.91}).\ Following (\ref{3.92}), we assume that $\psi ^{\ast }\left( 
\mathbf{x},s\right) =\psi ^{0}\left( \mathbf{x},s\right) $ for $\left( 
\mathbf{x},s\right) \in \left( \partial \Omega \setminus \Gamma \right)
\times \left[ \underline{s},\overline{s}\right] .$Hence, (\ref{3.8}) and (%
\ref{3.91}) hold for functions $q^{\ast },\psi ^{\ast }.$ Setting in (\ref%
{3.8}) $s=\overline{s},$ we obtain 
\begin{equation}
\begin{split}
\Delta q^{\ast }+2\overline{s}^{2}\nabla q^{\ast }\nabla V^{\ast }+2%
\overline{s}\left( \nabla V^{\ast }\right) ^{2}& =0,~~\mathbf{x}\in \Omega ,
\\
q^{\ast }\mid _{\partial \Omega }& =\psi ^{\ast }\left( \mathbf{x},\bar{s}%
\right) ,~\mathbf{x}\in \partial \Omega .
\end{split}
\label{4.18**}
\end{equation}%
Next, truncating the second term in each of the asymptotics (\ref{2.B}) and (%
\ref{2C}), we obtain that there exists a function $p^{\ast }\left( \mathbf{x}%
\right) \in C^{2+\alpha }\left( \overline{\Omega }\right) $ such that 
\begin{equation}
\begin{split}
V^{\ast }\left( \mathbf{x},\overline{s}\right) & \approx \frac{p^{\ast
}\left( \mathbf{x}\right) }{\overline{s}},\text{ }\overline{s}\rightarrow
\infty , \\
q^{\ast }\left( \mathbf{x},\overline{s}\right) & =\partial _{\overline{s}%
}V^{\ast }\left( \mathbf{x},\overline{s}\right) \approx -\frac{p^{\ast
}\left( \mathbf{x}\right) }{\overline{s}^{2}},\text{ }\overline{s}%
\rightarrow \infty .
\end{split}
\label{asymptotic}
\end{equation}%
Substituting formulae (\ref{asymptotic}) into (\ref{4.18**}), we obtain the
following approximate Dirichlet boundary value problem for the function $%
p^{\ast }\left( x\right) $ 
\begin{equation}
\Delta p^{\ast }=0\text{ in }\Omega ,\text{ }p^{\ast }\in C^{2+\alpha
}\left( \overline{\Omega }\right) ,  \label{2.917}
\end{equation}%
\begin{equation}
p^{\ast }|_{\partial \Omega }=-\overline{s}^{2}\psi ^{\ast }\left( \mathbf{x}%
,\overline{s}\right) .  \label{2.918}
\end{equation}%
Thus, using (\ref{4.17*}) and (\ref{asymptotic}), we obtain the following
approximate mathematical model.

\textbf{Approximate mathematical model.}

\emph{We assume that there exists a function }$p^{\ast }\left( x\right) \in
C^{2+\alpha }\left( \overline{\Omega }\right) $\emph{\ such that the exact
tail function }$V^{\ast }\left( \mathbf{x},s\right) $\emph{\ has the form} 
\begin{equation}
V^{\ast }\left( \mathbf{x},s\right) =\frac{p^{\ast }\left( \mathbf{x}\right) 
}{s}=\frac{\ln w^{\ast }\left( \mathbf{x},s\right) }{s^{2}},\text{ }\forall
s\geq \overline{s},\text{ }  \label{2.914}
\end{equation}%
\emph{and the function }$q^{\ast }\left( \mathbf{x},\overline{s}\right) $%
\emph{\ is}%
\begin{equation*}
q^{\ast }\left( \mathbf{x},\overline{s}\right) =-\frac{p^{\ast }\left( 
\mathbf{x}\right) }{\overline{s}^{2}}\text{.}
\end{equation*}

Because of (\ref{2.917}), (\ref{2.918}) and (\ref{2.914}), we set for the
first tail 
\begin{equation}
V_{1,0}\left( \mathbf{x}\right) =\frac{p\left( \mathbf{x}\right) }{\overline{%
s}},  \label{2.923}
\end{equation}%
where the function $p(\mathbf{x})$ is the solution of the following
Dirichlet boundary value problem 
\begin{equation}
\Delta p=0\text{ in }\Omega ,\text{ }p\in C^{2+\alpha }\left( \overline{%
\Omega }\right) ,  \label{2.924}
\end{equation}%
\begin{equation}
p|_{\partial \Omega }=-\overline{s}^{2}\psi \left( \mathbf{x},\overline{s}%
\right) .  \label{2.925}
\end{equation}%
We point out that we calculate $V_{1,0}\left( \mathbf{x}\right) $ without
any advanced knowledge of a small neighborhood of the exact coefficient $%
\varepsilon _{r}^{\ast }(\mathbf{x}).$ Using (\ref{asymptotic})-(\ref{2.925}%
) and Schauder theorem \cite{LU}, we obtain 
\begin{equation}
\left\Vert V_{1,0}\left( \mathbf{x}\right) -V^{\ast }\left( \mathbf{x},%
\overline{s}\right) \right\Vert _{C^{2+\alpha }\left( \overline{\Omega }%
\right) }\leq C\overline{s}\left\Vert \psi ^{\ast }\left( \mathbf{x},%
\overline{s}\right) -\psi \left( \mathbf{x},\overline{s}\right) \right\Vert
_{C^{2+\alpha }\left( \partial \Omega \right) },  \label{2.926}
\end{equation}%
where the number $C=C\left( \Omega \right) >0$ depends only from the domain $%
\Omega .$ Hence, the error in the calculation of $V_{1,0}\left( \mathbf{x}%
\right) $ depends only on the error in the boundary data $\psi \left( 
\mathbf{x},\overline{s}\right) .$ On the other hand, since the boundary
function $\psi \left( \mathbf{x},s\right) $ is generated by the function $g(%
\mathbf{x},t)$ in (\ref{2.5}), then the error in $\psi \left( \mathbf{x},%
\overline{s}\right) $ is generated by the error in measurements. The
estimate (\ref{2.926}) is one of elements of the proof of the approximate
global convergence theorem for this numerical method, see Theorem 2.9.4 in 
\cite{BK} and Theorem 5.1 in \cite{BKJIIP12}. Although a good approximation
for the exact solution $\varepsilon _{r}^{\ast }(\mathbf{x})$ can be derived
from the function $V_{1,0}\left( \mathbf{x}\right) ,$ we have observed
computationally that better approximations are delivered via iterations
described below in sections \ref{sec:6.1}, \ref{sec:6.2}.

\subsection{Discretization with respect to the pseudo-frequency}

\label{sec:4.3}

To approximate both functions $q$ and $V$ using (\ref{3.8}) and (\ref{3.91}%
), we consider the layer stripping procedure with respect to $s$. We divide
the interval $\left[ \underline{s},\overline{s}\right] $ into $N$ small
subintervals with the uniform step size $h=s_{n-1}-s_{n}$. Here, $\underline{%
s}=s_{N}<s_{N-1}<...<s_{0}=\overline{s}.$ We approximate the function $q(%
\mathbf{x},s)$ as a piecewise constant function with respect to $s,$ i.e. we
assume that $q(\mathbf{x},s)=q_{n}(\mathbf{x})$ for $s\in \left[
s_{n},s_{n-1}\right) .$ Hence, using (\ref{3.7}), we approximate the
function $v\left( \mathbf{x},s_{n}\right) $ as 
\begin{equation}
v\left( \mathbf{x},s_{n}\right) =-h\sum\limits_{j=0}^{n}q_{j}\left( \mathbf{x%
}\right) +V\left( \mathbf{x},\overline{s}\right) ,q_{0}\left( \mathbf{x}%
\right) :\equiv 0.  \label{4.20}
\end{equation}%
To obtain a sequence of Dirichlet boundary value problems for elliptic PDEs
for functions $q_{n}(\mathbf{x}),$ we introduce the $s-$dependent Carleman
Weight Function (CWF) $\mathcal{C}_{n,\mu }\left( s\right) =\exp \left[ \mu
\left( s-s_{n-1}\right) \right] ,$ where $\mu >>1$ is a large parameter. In
our numerical studies we take $\mu =20$. This function mitigates the
influence of the nonlinear term in the resulting integral-differential
equations on every pseudo-frequency interval $\left( s_{n},s_{n-1}\right) $.

Multiply both sides of equation (\ref{3.8}) by $\mathcal{C}_{n,\mu }\left(
s\right) $ and integrate with respect to $s\in \left( s_{n},s_{n-1}\right) .$
We obtain 
\begin{equation}
\begin{split}
&\Delta q_{n}-A_{1,n}\left( h\sum\limits_{j=0}^{n-1}\nabla q_{j}-\nabla
V_{n}\right) \nabla q_{n} \\
&=B_{n}\left( \nabla q_{n}\right) ^{2}-A_{2,n}h^{2}\left(
\sum\limits_{j=0}^{n-1}\nabla q_{j}\right) ^{2} +2A_{2,n}\nabla V_{n}\left(
h\sum\limits_{j=0}^{n-1}\nabla q_{j}\right) -A_{2,n}\left( \nabla
V_{n}\right) ^{2}, \\
&q_{n}\left( \mathbf{x}\right)  \mid _{\partial \Omega }=\psi _{n}(\mathbf{x}%
):=\frac{1}{h}\int\limits_{s_{n}}^{s_{n-1}}\psi \left( \mathbf{x},s\right)
ds,\text{ }n=1,...,N.
\end{split}
\label{4.21}
\end{equation}%
Here $V_{n}\left( \mathbf{x}\right) $ is such an approximation of the tail
function $V\left( \mathbf{x}\right) $ which corresponds to the function $%
q_{n}\left( \mathbf{x}\right) $ (section \ref{sec:6.1}). Numbers $%
A_{1,n},A_{2,n},B_{n}$ are computed explicitly. Furthermore, $B_{n}=O\left(
1/\mu \right) ,\mu \rightarrow \infty .$ For this reason we ignore the
nonlinear term in (\ref{4.21}), thus setting%
\begin{equation}
B_{n}\left( \nabla q_{n}\right) ^{2}:=0.  \label{4.22}
\end{equation}%
Note that (\ref{4.22}) is not a linearization, since (\ref{4.21}) contains
products $\nabla q_{j}\nabla q_{i}$ and also because the tail function $%
V_{n} $ depends nonlinearly on functions $q_{j},$ see (\ref{2.A}) and step
6 in section \ref{sec:6.1}.

\section{A Finite Element Method for the Reconstruction of $\protect%
\varepsilon _{r}\left( \mathbf{x}\right) $}

\label{sec:5}

In this section we explain how we compute functions $\varepsilon _{rn}(%
\mathbf{x})$ on every pseudo-frequency interval $\left( s_{n},s_{n-1}\right) 
$ using the FEM. Once the functions $q_{j},j=1,...n$ along with the function 
$V_{n}$ in (\ref{4.21}) are calculated, we compute the function $v_{n}\left( 
\mathbf{x}\right) $ using the direct analog of (\ref{4.20}), 
\begin{equation*}
v_{n}\left( \mathbf{x}\right) = -h\sum\limits_{j=0}^{n}q_{j}\left( \mathbf{x}%
\right) +V_{n}\left( \mathbf{x}\right) ,\text{ }x\in \Omega .
\end{equation*}%
Using (\ref{transform}), we set 
\begin{equation}
w_{n}\left( \mathbf{x}\right) =\exp \left[ s_{n}^{2}v_{n}\left( \mathbf{x}%
\right) \right] .  \label{3.104}
\end{equation}
To find the function $\varepsilon _{rn}\left( \mathbf{x}\right) ,$ we note
that the function $w_{n}\left( \mathbf{x},s_{n}\right) $ is the solution of
the following analog of the problem (\ref{2.7}), (\ref{2.8})%
\begin{equation}
\Delta w_{n}-s_{n}^{2}\varepsilon _{rn}\left( \mathbf{x}\right) w_{n}=0\text{
in }\Omega ,  \label{3.105}
\end{equation}%
\begin{equation}
\partial _{n}w_{n}\mid _{\partial \Omega }=f_{n}\left( \mathbf{x}\right) ,
\label{3.106}
\end{equation}%
where 
\begin{equation}
f_{n}\left( \mathbf{x}\right) =\partial _{n}\exp \left[ s_{n}^{2}v_{n}\left( 
\mathbf{x}\right) \right] \text{ for }\mathbf{x}\in \partial \Omega .
\label{3.107}
\end{equation}%
To compute the function $\varepsilon _{rn}\left( \mathbf{x}\right) $ from (%
\ref{3.105}), (\ref{3.106}) and (\ref{3.107}), we apply a version of the FEM
as described below in sections \ref{sec:5.1}, \ref{sec:5.2}.

\subsection{Spaces of finite elements}

\label{sec:5.1}

Following \cite{Johnson} we discretize in computations our bounded domain $%
\Omega \subset \mathbb{R}^{3}$ by an unstructured tetrahedral mesh $T$ using
non-overlapping tetrahedral elements $K\in \mathbb{R}^{3}$. The elements $K$
are such that $T=\{K_{1},...,K_{m}\}$, where $m$ is the total number of elements
in $\Omega $, and 
\begin{equation*}
\Omega =\cup _{K\in T}K=K_{1}\cup K_{2}...\cup K_{m}.
\end{equation*}%
We associate with the mesh $T$ the mesh function $h=h(\mathbf{x})$ as a
piecewise-constant function such that 
\begin{equation*}
h(\mathbf{x})=h_{K},~\forall K\in T,
\end{equation*}%
where $h_{K}$ is the diameter of $K$ which we define as the longest side of $%
K$. We impose the following shape regularity assumption of the mesh $T$ for
every element $K\in T$ 
\begin{equation}
a_{1}\leq h_{K}\leq r^{\prime }a_{2},\quad a_{1},a_{2}=const.>0,  \label{5.1}
\end{equation}%
where $r^{\prime }$ is the radius of the maximal sphere contained in the
element $K$.

Define the set of polynomials $P_{r}(K)$ as 
\begin{equation}
P_{r}(K)=\big\{v:v(x,y,z)=\sum_{0\leq i+j+l\leq
r}c_{ijl}x^{i}y^{j}z^{l},(x,y,z)\in K,c_{ijl}\in \mathbb{R},~\forall K\in T%
\big\}.  \label{polynom}
\end{equation}%
We introduce now the finite element space $V_{h}$ as 
\begin{equation*}
V_{h}=\big\{v(x)\in H^{1}\left( \Omega \right) :v\in C(\Omega ),~v|_{K}\in
P_{1}(K)~\forall K\in T\big\},
\end{equation*}%
where $P_{1}(K)$ denotes the set of linear functions on $K$ defined by (\ref%
{polynom}) for $r=1$. Hence, the finite element space $V_{h}$ consists of
continuous piecewise linear functions in $\Omega $. To approximate functions 
$\varepsilon _{rn},$ we introduce the space of piecewise constant functions $%
C_{h}$, 
\begin{equation}
C_{h}:=\{u\in L_{2}(\Omega ):u|_{K}\in P_{0}(K),\forall K\in T\},  \notag
\end{equation}%
where $P_{0}(K)$ is the piecewise constant function on $K$ defined by (\ref%
{polynom}) for $r=0$.

\subsection{A finite element method}

\label{sec:5.2}

To compute the function $\varepsilon _{rn}$ from (\ref{3.105}), we formulate
the finite element method for the problem (\ref{3.106})-(\ref{3.107}) as:
Find the function $\varepsilon _{rn}\in C_{h}$ for the known function $%
w_{n}\in V_{h}$ such that 
\begin{equation}
(\varepsilon _{rn}w_{n},v)=-\frac{1}{s_{n}^{2}}(\nabla w_{n},\nabla v)+\frac{%
1}{s_{n}^{2}}(f_{n},v)_{\partial \Omega },\forall v\in V_{h},
\label{3.107_1}
\end{equation}%
where $(\cdot ,\cdot )$ is the scalar product in $L_{2}\left( \Omega \right) 
$.

We expand $w_{n}$ in terms of the standard continuous piecewise linear
functions $\{\varphi _{l}\}_{l=1}^{P}$ in the space $V_{h}$ as 
\begin{equation}
w_{n}(\mathbf{x})=\sum_{l=1}^{P}w_{n,l}\varphi _{l}(\mathbf{x}),
\label{3.200}
\end{equation}%
where $w_{n,l}$ denote the nodal values of the function $w_{n}$ at the nodes 
$l$ of the elements $K$ in the mesh $T$. We can determine $w_{n,l}$ by
knowing already computed functions $v_{n,l}$ using the following analog of (%
\ref{3.104}) 
\begin{equation*}
w_{n}\left( \mathbf{x}\right) =\exp \left[ s_{n}^{2}v_{n,l}\left( \mathbf{x}%
\right) \right] ,\forall \mathbf{x}\in \Omega .
\end{equation*}%
Substitute (\ref{3.200}) into (\ref{3.107_1}) and choose $v(\mathbf{x}%
)=\varphi _{j}(\mathbf{x}).$ Then we obtain the following linear algebraic
system of equations 
\begin{equation}
\sum_{l,j=1}^{P}\varepsilon _{rn,l}(w_{n,l}\varphi _{l},\varphi _{j})=-\frac{%
1}{s_{n}^{2}}\sum_{l,j=1}^{P}w_{n,l}(\nabla \varphi _{l},\nabla \varphi
_{j})+\frac{1}{s_{n}^{2}}\sum_{j=1}^{P}\left[ f_{n},\varphi _{j}\right] ,
\label{3.108}
\end{equation}%
where $\left[ \cdot ,\cdot \right] $ is the scalar product in $L_{2}\left(
\partial \Omega \right) .$ The system (\ref{3.108}) can be rewritten in the
matrix form for the unknown vector $\varepsilon _{rn}=\left\{ \varepsilon
_{rn,l}\right\} _{l=1}^{P}$ and known vector $w_{n}=\left\{ w_{n,l}\right\}
_{l=1}^{P}$ as 
\begin{equation}
M\varepsilon _{rn}=-\frac{1}{s_{n}^{2}}Gw_{n}+\frac{1}{s_{n}^{2}}F.
\label{3.108_1}
\end{equation}%
Here $M$ is the block mass matrix in space, $G$ is the stiffness matrix
corresponding to the term containing $(\nabla \varphi _{l},\nabla \varphi
_{j})$ in (\ref{3.108}) and $F$ is the load vector. At the element $K$ the
matrix entries in (\ref{3.108_1}) are explicitly given by: 
\begin{equation*}
M_{l,j}^{K}=(w_{n,l}~\varphi _{l},\varphi _{j})_{K},G_{l,j}^{K}=(\nabla
\varphi _{l},\nabla \varphi _{j})_{K},F_{n,j}^{K}=(f_{n},\varphi _{j})_{K}.
\end{equation*}

To obtain an explicit scheme for the computation of coefficients $%
\varepsilon _{rn}$, we approximate the matrix $M$ by the lumped mass matrix $%
M^{L}$ in space, i.e., the diagonal approximation is obtained by taking the
row sum of $M$ \cite{BK}. We obtain 
\begin{equation}
\varepsilon _{rn}=-\frac{1}{s_{n}^{2}}(M^{L})^{-1}Gw_{n}+\frac{1}{s_{n}^{2}}%
(M^{L})^{-1}F.  \label{3.109}
\end{equation}%
Note that for the case of linear Lagrange elements which are used in our
computations in section \ref{sec:8} we have $M=M^{L}$. Thus, the lumping
procedure does not include approximation errors in this case.

\section{The Approximately Globally Convergent Algorithm}

\label{sec:6}

We present now our algorithm for the numerical solution of equations (\ref%
{4.21}) and computing the functions $\varepsilon _{rn}$ using the equation(%
\ref{3.109}). In this algorithm the index $i$ denotes the number of inner
iterations inside every pseudo-frequency interval $\left(
s_{n},s_{n-1}\right) $ when we update tails.

\subsection{The algorithm}

\label{sec:6.1}

\begin{description}
\item[Step 0] Set $q_{0}=0$. Compute the initial tail function $V_{1,0}(%
\mathbf{x},\overline{s})\in C^{2+\alpha }(\overline{\Omega })$ as in (\ref%
{2.923})-(\ref{2.925}).

\item[Step 1] Here we describe iterations which update tails inside every
pseudo-frequency interval $\left( s_{n},s_{n-1}\right) $. Let $n\geq 1,i\geq
1.$ Suppose that functions $q_{j},j=1,...,n-1,V_{n,i-1}$ are computed. Solve
the Dirichlet boundary value problem for the function $q_{n,i}\left( \mathbf{%
x}\right) \in C^{2+\alpha }\left( \overline{\Omega }\right) ,$ 
\begin{equation}
\begin{split}
\Delta q_{n,i}-A_{1n}& \left( h\sum\limits_{j=1}^{n-1}\nabla q_{j}\right)
\cdot \nabla q_{n,i}+A_{1n}\nabla q_{n,i}\cdot \nabla V_{n,i-1}= \\
& -A_{2n}h^{2}\left( \sum\limits_{j=1}^{n-1}\nabla q_{j}\right)
^{2}+2A_{2n}\nabla V_{n,i-1}\cdot \left( h\sum\limits_{j=1}^{n-1}\nabla
q_{j}\right) -A_{2n}\left( \nabla V_{n,i-1}\right) ^{2}, \\
q_{n,i}\left( \mathbf{x}\right) & =\psi _{n}\left( \mathbf{x}\right) ,\qquad 
\mathbf{x}\in \partial \Omega .
\end{split}
\label{7.102}
\end{equation}

\item[Step 2] Compute functions $v_{n,i}\left( \mathbf{x}\right) $ and $%
w_{n,i}\left( \mathbf{x}\right) ,$ 
\begin{equation*}
v_{n,i}\left( \mathbf{x}\right) =-hq_{n,i}\left( \mathbf{x}\right)
-h\sum\limits_{j=0}^{n-1}q_{j}\left( \mathbf{x}\right) +V_{n,i}\left( 
\mathbf{x}\right) ,
\end{equation*}%
\begin{equation*}
w_{n,i}\left( \mathbf{x}\right) =\exp \left[ s_{n}^{2}v_{n,i}\left( \mathbf{x%
}\right) \right] .
\end{equation*}

\item[Step 3] Compute the function $\overline{\varepsilon }_{r,n,i}\in C_{h}$
via backwards calculations, using the finite element formulation of equation
(\ref{3.109}) as 
\begin{equation*}
\overline{\varepsilon }_{rn,i}\left( \mathbf{x}\right) =-\frac{1}{s_{n}^{2}}%
(M^{L})^{-1}Gw_{n,i}+\frac{1}{s_{n}^{2}}(M^{L})^{-1}F.
\end{equation*}%
Since by (\ref{2.20}) we should have $\varepsilon _{r}\left( \mathbf{x}%
\right) \geq 1,\forall \mathbf{x}\in \mathbb{R}^{3},$ and also since we need
to extend the function $\overline{\varepsilon }_{r,n,i}\left( \mathbf{x}%
\right) $ outside of the domain $\Omega $ by unity, we set 
\begin{equation}
\varepsilon _{rn,i}\left( \mathbf{x}\right) =\left\{ 
\begin{array}{c}
\overline{\varepsilon }_{rn,i}\left( \mathbf{x}\right) \text{ if }\overline{%
\varepsilon }_{r,n,i}\left( \mathbf{x}\right) \geq 1, \\ 
1\text{ if either }\overline{\varepsilon }_{rn,i}\left( \mathbf{x}\right) <1,%
\text{ or }\mathbf{x}\in \mathbb{R}^{3}\diagdown \Omega .%
\end{array}%
\right.  \label{7.101}
\end{equation}

\item[Step 4] Solve the forward problem (\ref{2.1})-(\ref{2.2}) with ${%
\varepsilon _{r}}(\mathbf{x}):=\varepsilon _{rn,i}\left( \mathbf{x}\right) $
and compute the Laplace transform (\ref{2.6}) for $s= s_n$. We
obtain the function $w_{n,i}\left( \mathbf{x}, s_n\right) $.

\item[Step 5] Update the tail function as%
\begin{equation}
V_{n,i}(\mathbf{x})=\frac{\ln w_{n,i}\left( \mathbf{x}, s_n\right) }{{s_n}^{2}}.  \label{7.103}
\end{equation}%
Continue inner iterations with respect to $i$ until the stopping criterion
of Step 1 of section \ref{sec:6.2} is met at $i=m_{n}$.

\item[Step 6] Set for the pseudo-frequency interval $[s_{n},s_{n-1})$ 
\begin{equation}
q_{n}(\mathbf{x}):=q_{n,m_{n}}(\mathbf{x}),{\varepsilon }_{rn}(\mathbf{x}):={%
\varepsilon }_{rn,m_{n}}(\mathbf{x}),V_{n+1,0}\left( \mathbf{x}\right) :=%
\frac{\ln w_{n,m_{n}}\left( \mathbf{x},s_n\right) }{{s_n}^{2}%
}:=V_{n}\left( \mathbf{x}\right) .  \label{7.104}
\end{equation}

\item[Step 7] If either the stopping criterion with respect to $n$ of Step 4
of section \ref{sec:6.2} is met, or $n=N,$ then set the resulting function ${%
\varepsilon }_{rn}(\mathbf{x})$ as the solution of our CIP. Otherwise, set $%
n:=n+1$ and go to Step 1.
\end{description}

\subsection{The stopping criterion}

\label{sec:6.2}

When testing the algorithm of section \ref{sec:6.1} on experimental data, we
have developed a reliable stopping criterion for iterations $(n,i)$ in this
algorithm. On every pseudo-frequency interval $\left( s_{n},s_{n-1}\right) $
we define \textquotedblleft first norms\textquotedblright\ $D_{n,0}$ as 
\begin{equation}
D_{n,0}=||{V}_{n,0}-\tilde{V}_{n}||_{L_{2}(\Omega )}.  \label{first_norms}
\end{equation}%
In (\ref{first_norms}) the function ${V}_{n,0}$ is the computed tail
functions at the inner iteration $i=0$ as in (\ref{7.104}). Functions $%
\tilde{V}_{n}$ in (\ref{first_norms}) are obtained from the known measured
function $g(\mathbf{x},t)$ in (\ref{2.5}) as 
\begin{equation}
\tilde{V}_{n}\left( \mathbf{x}\right) =\frac{\ln W(\mathbf{x},{s_{n}})}{%
s_{n}^{2}},  \label{obstail}
\end{equation}%
where $W(\mathbf{x},s_{n})$ is the Laplace transform of the function $g(%
\mathbf{x},t)$ at $s=s_{n}$.

We have observed that computed \textquotedblleft first
norms\textquotedblright\ $D_{n,0}$ always achieve only one minimum at a
certain $n=\overline{n}$, where the number $\overline{n}$ depends on the
specific set of experimental data.\ Furthermore, in non-blind cases of
non-metallic targets, the corresponding values of $\max_{\overline{\Omega }}{%
\varepsilon _{r\overline{n},0}}(\mathbf{x})$ were in a good agreement with a
priori known ones. However, in the cases of non-blind metallic targets we
have observed that $5\leq \max_{\overline{\Omega }}{\varepsilon _{r\overline{%
n},0}}(\mathbf{x})\leq 10.$ This contradicts with (\ref{2.51}). Therefore,
we have developed the following stopping criterion which consists of four steps.

\textbf{The Stopping Criterion}

 The first step in our criterion is for stopping inner iterations with
 respect to $i$ in step 5 of section \ref{sec:6.1}. As to Steps 2-4,
 they are for stopping outer iterations with respect to $n$ (Step 7 in
 section \ref{sec:6.1}). First, we define numbers $B_{n,i}$ and
 $D_{n,i}$ as%
\begin{equation*}
B_{n,i}=\frac{||{\varepsilon _{rn,i}}-{\varepsilon _{rn,i-1}}%
||_{L_{2}(\Omega )}}{||{\varepsilon _{rn,i-1}}||_{L_{2}(\Omega )}},
\end{equation*}%
\begin{equation}
D_{n,i}=||{V}_{n,i}-\tilde{V}_{n}||_{L_{2}(\Omega )},  \label{7.6_1}
\end{equation}%
In (\ref{7.6_1}) functions ${V}_{n,i}$ are computed tail functions
corresponding to ${\varepsilon _{rn,i}}$ (step 6 in section \ref{sec:6.1})
and functions $\tilde{V}_{n}=\tilde{V}_{n}(\mathbf{x},s_{n})$ are calculated
using (\ref{obstail}).

\begin{itemize}
\item \textbf{Step 1.} Iterate with respect to $i$ and stop iterations at $%
i=m_{n}\geq 1$ such that%
\begin{equation}
\text{either }B_{n,i}\geq B_{n,i-1}\text{ or }B_{n,i}\leq \eta,
\label{7.5}
\end{equation}%
or 
\begin{equation}
\text{either }\text{{}}D{_{n,i}}\geq D{_{n,i-1}}\text{ or }D{_{n,i}}\leq \eta,  \label{7.5_1}
\end{equation}
where $\eta=10^{-6}$ is a chosen tolerance.
%Therefore, first inequalities in (\ref{7.5}), (\ref{7.5_1}) mean that
%numbers $B_{n,i},D_{n,i}$ start to increase with respect to $i$.

\item \textbf{Step 2. }For every $n$\textbf{\ }compute \textquotedblleft
final norms\textquotedblright\ $D_{n,m_{n}}$ as 
\begin{equation}
D_{n,m_{n}}=||{V}_{n+1,0}-\tilde{V}_{n}||_{L_{2}(\Omega )}.
\label{final_norms}
\end{equation}%
In (\ref{final_norms}) functions ${V}_{n+1,0}\left( x\right) $ are computed
as in (\ref{7.104}).

\item \textbf{Step 3}. Compute the number $\overline{N}$ of the pseudo
frequency interval such that the first norms $D_{n,0}$ in (\ref{first_norms})
  achieve its first minimum with respect to $n$ and get corresponding  
$\varepsilon _{r\overline{N},0}( \mathbf{x})$ on this interval.
 Compute the number $\overline{M}$ of the pseudo frequency interval
 such that the final norms $D_{n,m_n}$ in (\ref{final_norms})
 achieve its first minimum or they are stabilized with respect to $n$,
 and get corresponding $\varepsilon _{r\overline{M},0}( \mathbf{x})$ on this
 interval.
Next, compute the number $\tilde{\varepsilon}_{r},$
\begin{equation}
\tilde{\varepsilon}_{r}=\left\{ 
\begin{array}{cc}
\max_{\overline{\Omega }}\varepsilon_{r\overline{M},0}\left( \mathbf{x}\right), & \text{  if } \overline{M} < \overline{N},\\
\max_{\overline{\Omega }}\varepsilon_{r\overline{N},0}\left( \mathbf{x}\right), & \text{  if } \overline{M} \geq \overline{N}.
\label{7.7}
\end{array}
\right.
\end{equation}

\item \textbf{Step 4}. If $\tilde{\varepsilon}_{r}<5$ or $\tilde{\varepsilon}_{r}>10,$ then take the final reconstructed value of the
refractive index $n=\sqrt{\tilde{\varepsilon}_{r}}.$ As the computed
function $\varepsilon_{r}\left( \mathbf{x}\right),$ take 
%$\varepsilon_{r,comp}\left( \mathbf{x}\right) =\varepsilon _{r\overline{N},0}\left( \mathbf{x}\right) $ and st%op iterations. 
\begin{equation}
\varepsilon
_{r,comp}\left( \mathbf{x}\right)=\left\{ 
\begin{array}{cc}
\varepsilon _{r\overline{M},0}\left( \mathbf{x}\right), & \text{  if } \overline{M} < \overline{N},\\
\varepsilon _{r\overline{N},0}\left( \mathbf{x}\right), &  \text{  if } \overline{M} \geq \overline{N}.
\label{compeps}
\end{array}
\right.
\end{equation}
 and stop iterations. 
However, if $5\leq \tilde{\varepsilon}_{r}\leq 10,$ then continue iterations and compute
the number $\widetilde{N}\in \left( \overline{N}+1,N\right] $ of the pseudo
frequency interval such that the global minimum with respect to $n$ of
final norms $D_{n,m_{n}}$ in (\ref{final_norms}) is achieved. Then,
similarly with (\ref{7.7}), compute the number $\varepsilon _{r\widetilde{N}%
},$ 
\begin{equation}
\varepsilon _{r\widetilde{N}}=\max_{\overline{\Omega }}\varepsilon _{r%
\widetilde{N},0}\left( \mathbf{x}\right)  \label{7.8}
\end{equation}%
and take $n=\sqrt{\varepsilon _{r\widetilde{N}}}$ as the final reconstructed
value of the refractive index. Also, take the function $\varepsilon
_{r,comp}\left( \mathbf{x}\right) =\varepsilon _{r\widetilde{N},0}\left( \mathbf{x}\right) $ as the computed coefficient $%
\varepsilon _{r}\left( \mathbf{x}\right) $ and stop iterations.
\end{itemize}

\begin{figure}[tbp]
\begin{center}
 \subfloat[a) Test 1]{\includegraphics[scale=0.47,clip=]{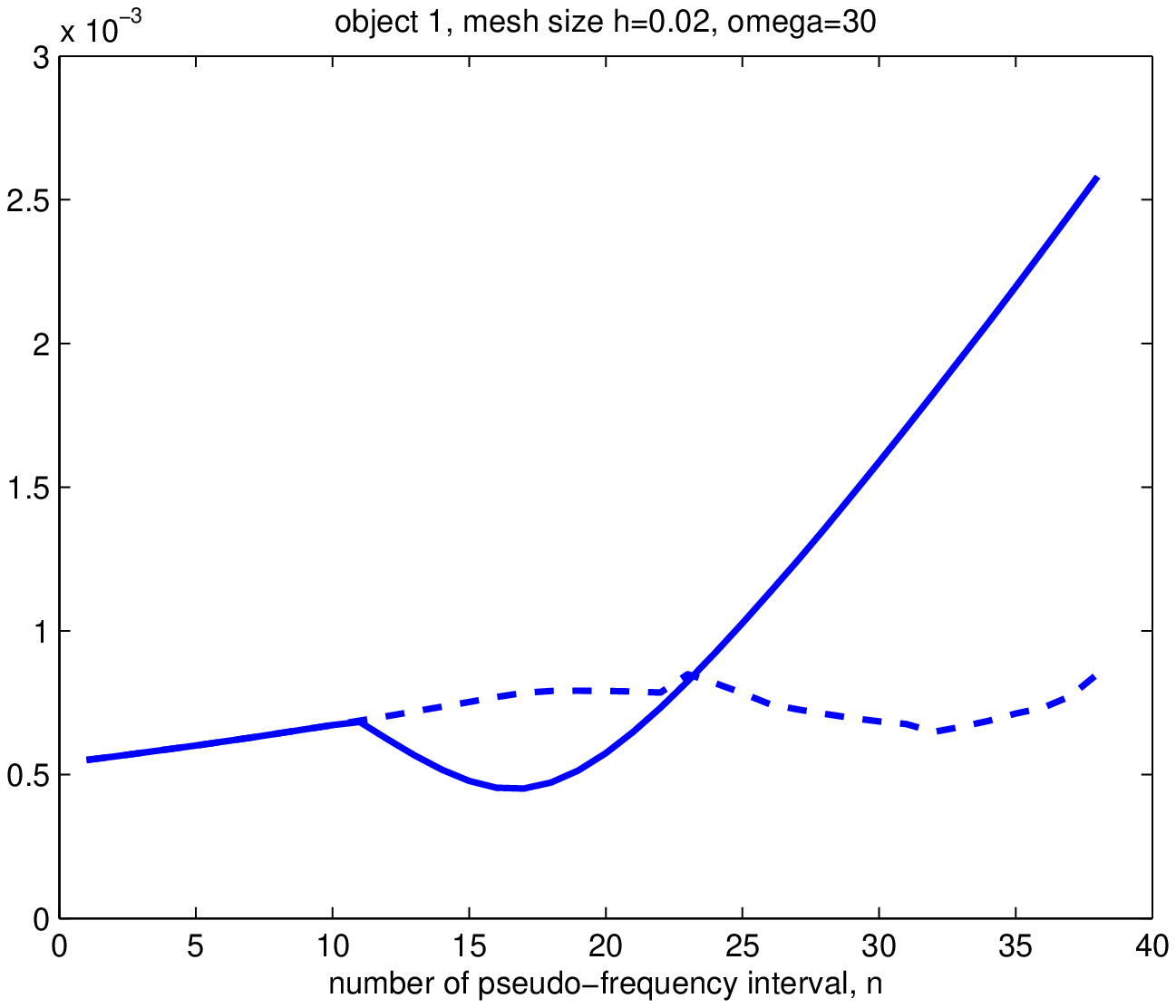}} 
 \subfloat[b) Test 2]{\includegraphics[scale=0.47,clip=]{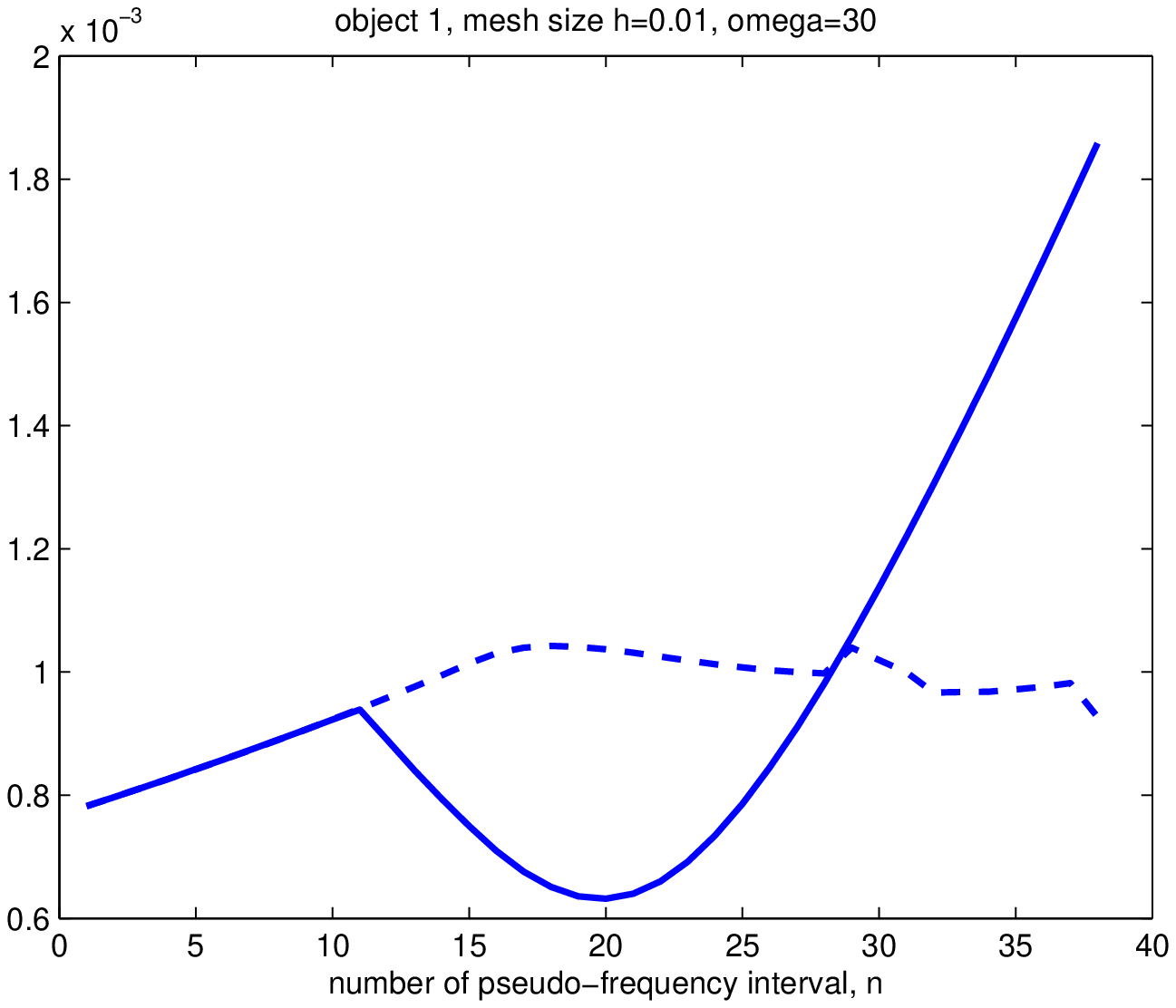}}
\end{center}
\caption{\protect\small \emph{ Behaviour of norms $D_{n,m_n}$ (solid curve) and $D_{n,m_n}$  (dashed curve) for object 1.}}
\label{fig:finalnorms}
\end{figure}

We have observed in all our computations that conditions of our stopping
criterion are always achieved.\ More precisely, one of conditions (\ref{7.5}%
), (\ref{7.5_1}) is always achieved for iterations with respect to $i$ and
the minimal values mentioned in Steps 3 and 4 are always achieved. 
Figure \ref{fig:finalnorms}  displays a typical $n-$dependence of sequences $D_{n,0\text{ }}$and $D_{n,m_{n}}.$
% 
% 
% \begin{figure}[tbp]
% \begin{center}
% \begin{tabular}{cc}
% {\includegraphics[scale=0.5,clip=]{F3a.eps}} &
%  {\includegraphics[scale=0.5,clip=]{F3b.eps}}\\
% a)  Test 1 & b)  Test 2 \\
% \end{tabular}
% \end{center}
% \caption{\protect\small \emph{ Behaviour of norms $D_{n,m_n}$ (solid curve) and $D_{n,m_n}$  (dashed curve) for object 1.}}
% \label{fig:finalnorms}
% \end{figure}

\section{Some Details of the Numerical Implementation}

\label{sec:7}

In this section we present some additional details of our numerical
implementation. Because of (\ref{2.51}), we define in all our tests the
upper value of the function $\varepsilon_{r}\left(\mathbf{x}\right)$ 
as $b=15,$  see (\ref{2.20}). Thus, we set lower and upper bounds for the
reconstructed function $\varepsilon_{r}(\mathbf{x})$ in $\Omega$ as 
\begin{equation}
M_{\varepsilon _{r}}=\{\varepsilon _{r}(\mathbf{x}):\varepsilon _{r}\left( 
\mathbf{x}\right) \in \left[ 1, 15\right] \}.  \label{4.30}
\end{equation}%
As to the lower bound, we ensure it via (\ref{7.101}). We ensure the upper
bound 15 similarly via truncating to 15 those values of $\varepsilon
_{r,comp}\left( \mathbf{x}\right) $ which exceed this number. To solve
Dirichlet boundary value problems (\ref{7.102}), we use FEM. We reconstruct
refractive indices rather than dielectric constants of material since they
can be directly measured.

To compare our computational results with directly measured refractive
indices $n=\sqrt{\varepsilon _{r}}$ of dielectric targets and with appearing
dielectric constants of metallic targets (see (\ref{2.51})), we consider
maximal values of computed functions $\varepsilon _{r,comp}\left( \mathbf{x}%
\right) $, 
\begin{equation}
\varepsilon _{r}^{\text{comp}}=\max_{\overline{\Omega }}\varepsilon
_{r,comp}\left( \mathbf{x}\right) ,n^{\text{comp}}=\sqrt{\varepsilon _{r}^{%
\text{comp}}},  \label{4.300}
\end{equation}
see Step 4 of section \ref{sec:6.2} for the definition of $\varepsilon
_{r,comp}\left( \mathbf{x}\right) .$ Using experimental data for non-blind
targets and comparing reconstruction results with cases of synthetic data,
we have found that our algorithm provided accurate results with the
following pseudo frequency interval, which we use in all our computations 
\begin{equation*}
s\in \lbrack 8,10],\underline{s}=8,\overline{s}=10,h=0.05.
\end{equation*}

\subsection{Computations of the forward problem}

\label{sec:7.1}

As it is clear from Step 4 of section \ref{sec:6.1}, we need to solve
the forward problem (\ref{2.1}), (\ref{2.2}) on each iterative step of
inner iterations to update the tail via (\ref{7.103}). Since it is
impossible to computationally solve equation (\ref{2.1}) in the
infinite space $\mathbb{R}^{3},$ we work with a truncated
domain. Namely, we choose the domain $G$ as
\begin{equation*}
G=\left\{ \mathbf{x=}(x,y,z)\in (-0.56,0.56)\times (-0.56,0.56)\times
(-0.16,0.1)\right\}. 
\end{equation*}
 We use the hybrid FEM/FDM method described in
\cite{BSA} and the software package WavES \cite{waves}. We split $G$
into two subdomains $G_{FEM}= \Omega $ and $G_{FDM}$ so that
$G=G_{FEM}\cup G_{FDM}$. We solve the
forward problem in $G$ and the inverse problem via the
algorithm of section \ref{sec:6.1} in $\Omega.$ The space mesh in
$G_{FEM}$ and in $G_{FDM}$ consists of tetrahedral and cubes,
respectively. Below
\begin{equation}
G_{FEM}=\Omega =\left\{ \mathbf{x=}(x,y,z)\in (-0.5,0.5)\times
(-0.5,0.5)\times (-0.1,0.04)\right\} .  \label{8.0}
\end{equation}
Since by (\ref{2.20}) $\varepsilon _{r}(\mathbf{x})=1$ in $G_{FDM},$
then it is computationally efficient to use FDM in $G_{FDM}$ and to
use FEM in $G_{FEM}=\Omega,$ as it is done in the hybrid method of
\cite{BSA}.

The front and back sides of the rectangular prism $G$ are $\{z=0.1\}$
and $\{z=-0.16\}$, respectively. The boundary of the domain $G$ is
$\partial G=\partial _{1}G\cup \partial _{2}G\cup \partial _{3}G.$
Here, $\partial _{1}G$ and $\partial _{2}G$ are, respectively, front
and back sides of the domain $G$, and $\partial _{3}G$ is the union of
left, right, top and bottom sides of this domain. The front side
$\Gamma $ of the rectangular prism $\Omega $ where the propagated
data $g\left( \mathbf{x},t\right) $ in (\ref{2.5})  are given, is
\begin{equation}
\Gamma =\{\mathbf{x}\in \partial \Omega :z=0.04\}  \label{8.01}
\end{equation}%

Now we describe the forward problem which is used in our computations. To
compute tail functions $V_{n,i}$ via Steps 4, 5 of the algorithm of section %
\ref{sec:6.1}, we computationally solve the following forward problem in our
tests: 
\begin{equation}
\begin{split}
\varepsilon _{r}\left( \mathbf{x}\right) u_{tt}-\Delta u& =0,~~~\mbox{in}%
~G\times (0,T), \\
u(\mathbf{x},0)& =0,~u_{t}(\mathbf{x},0)=0,~\mbox{in}~G, \\
\partial _{n}u& =f\left( t\right) ,~\mbox{on}~\partial _{1}G\times (0,t_{1}],
\\
\partial _{n}u& =-\partial _{t}u,~\mbox{on}~\partial _{1}G\times (t_{1},T),
\\
\partial _{n}u& =-\partial _{t}u,~\mbox{on}~\partial _{2}G\times (0,T), \\
\partial _{n}u& =0,~\mbox{on}~\partial _{3}G\times (0,T),
\end{split}
\label{8.2}
\end{equation}%
where $f(t)$ is the amplitude of the initialized plane wave, 
\begin{equation*}
f(t)=\sin \omega t,~0\leq t\leq t_{1}:=\frac{2\pi }{\omega }.
\end{equation*}%
We use $\omega =30$ and $T=1.2.$ We solve the problem (\ref{8.2}) using the
explicit scheme with the time step size $\tau =0.003,$ which satisfies the
CFL condition. 

\subsection{Two stages}

\label{sec:7.2}

Our reconstruction procedure is done in two stages described in this section.

\subsubsection{First stage}

\label{sec:7.2.1}

In the first stage we follow the algorithm of section \ref{sec:6.1}. We have
observed that this stage provides accurate locations of targets of
interest.\ It also provides accurate values of refractive indices $n=\sqrt{%
\varepsilon _{r\overline{N}}}$ of dielectric targets and large values of
appearing dielectric constants $\varepsilon _{r\widetilde{N}}$ for metallic
targets, see (\ref{7.7}) and (\ref{7.8}). However, the algorithm of section %
\ref{sec:6.1} does not reconstruct well sizes/shapes of targets. Thus, we
need a postprocessing procedure, which is done in the second stage.

\subsubsection{The second stage: postprocessing}

\label{sec:7.2.2}

Let $\varepsilon _{rn,i}\left( \mathbf{x}\right) $ be the function in (\ref%
{7.101}). Then we set 
\begin{equation}
\widetilde{\varepsilon }_{rn,i}(\mathbf{x})=\left\{ 
\begin{array}{ll}
\varepsilon _{rn,i}(\mathbf{x}) & \text{ if }\varepsilon _{rn,i}(\mathbf{x}%
)>0.5\max\limits_{\Omega }\varepsilon _{rn,i}(\mathbf{x}), \\ 
1, & \text{ otherwise. }%
\end{array}%
\right.  \label{7.9}
\end{equation}%
Next, we determine minimal $x_{\min },y_{\min }$ and maximal $x_{\max
},y_{\max }$ values in $x$ and $y$ directions, where the function $%
\widetilde{\varepsilon }_{rn,i}(\mathbf{x})>1.$ Next, we set%
\begin{equation*}
\varepsilon _{rn,i}\left( \mathbf{x}\right) :=%
\begin{cases}
\widetilde{\varepsilon }_{rn,i}(\mathbf{x}) & \text{ if }x\in \left[ x_{\min
},x_{\max }\right] ,y\in \left[ y_{\min },y_{\max }\right] , \\ 
1 & \text{ otherwise}%
\end{cases}%
\end{equation*}%
and proceed with Step 5 of the algorithm of section \ref{sec:6.1}. In this
second stage we perform the same number of iterations with respect to both
indices $n,i$ as ones of the first stage. We are concerned in the second
stage only with sized and shapes of targets, and we are not concerned with
values of $\varepsilon _{r}^{\text{comp}},n^{\text{comp}}.$ Rather, we take
these values from the first stage. Let $\widetilde{\varepsilon }_{r}\left( 
\mathbf{x}\right) $ be the function $\varepsilon _{r}\left( \mathbf{x}%
\right) $ obtained at the last iteration of the second stage. Then we form
the image of the target based on the function $\varepsilon _{r,image}\left( 
\mathbf{x}\right) ,$%
\begin{equation*}
\varepsilon _{r,image}\left( \mathbf{x}\right) =\left\{ 
\begin{array}{l}
\widetilde{\varepsilon }_{r}\left( \mathbf{x}\right) \text{ if }\widetilde{%
\varepsilon }_{r}\left( \mathbf{x}\right) \geq 0.9\max_{\overline{\Omega }}%
\widetilde{\varepsilon }_{r}\left( \mathbf{x}\right) , \\ 
1\text{ otherwise.}%
\end{array}%
\right.
\end{equation*}

\begin{table}
\begin{center}
\begin{tabular}{| c | p{4.7cm}  |  p{4.7cm} }
\hline
\textbf{ Object number} &  \textbf{Name of the object}  \\
\hline 
  1 & a piece of oak \\ 
  2 & a piece of pine \\
  3 &  a metallic sphere \\
  4 & a metallic cylinder\\
  5 & blind target \\
  6 & blind target  \\
  7 &   blind target  \\
  8 &   doll, air inside, blind target \\
 9 &    doll, metal inside, blind target \\
 10 &  doll, sand inside, blind target \\
 11 &  two metallic blind targets \\
\hline
\end{tabular}
\end{center}
\caption{\emph{Object names.}}
\label{tab:table1}
\end{table}

\section{Results}

\label{sec:8}

Goals of our computational studies are: (1) To differentiate between dielectric and metallic targets,
(2) To reconstruct refractive indices of dielectric targets and appearing
dielectric constants of metallic targets, (3) To image locations of targets,
their sizes and sometimes their shapes. It is more challenging to compute
sizes of targets in the $z-$direction (i.e. depth) than in $x,y$ directions.

\subsection{Three tests}

\label{sec:8.1}

To see how sensitive the algorithm is to $x,y$ sizes of the prism $\Omega $
as well as to the mesh step size $h_{\mathbf{x}}$ in computations of both
forward and inverse problems, we run the above numerical procedure for all
our targets for the following three tests:

Test 1. The domain $\Omega $ for the computation of the CIP is as in
(\ref{8.0}) and the mesh step size is $h_{\mathbf{x}}=0.02$. Recall
that the distance between neighboring positions of our detector on the
measurement plane $P_{m}$ is also 0.02.

Test 2. The domain $\Omega$ is as in (\ref{8.0}). But the mesh step size
here is $h_{\mathbf{x}}=0.01.$

Test 3. In this test we shrink the domain $\Omega $ in $x,y$
directions, while keeping the same mesh size $h_{\mathbf{x}}=0.02$ as
in Test 1. In this test
\begin{equation}\label{8.1}
G_{FEM}=\Omega =\left\{ \mathbf{x=}(x,y,z)\in (-0.2,0.2)\times
(-0.2,0.2)\times (-0.1,0.04)\right\} ,
\end{equation}
\begin{equation}
M_{\varepsilon _{r}}=\{\varepsilon _{r}(\mathbf{x}):\varepsilon _{r}\left( 
\mathbf{x}\right) \in \left[ 1,15\right] \}.  \label{8.20}
\end{equation}

\begin{figure}[tbp]
\begin{center}
\begin{tabular}{cc}
{\includegraphics[width=0.45\textwidth,clip=]{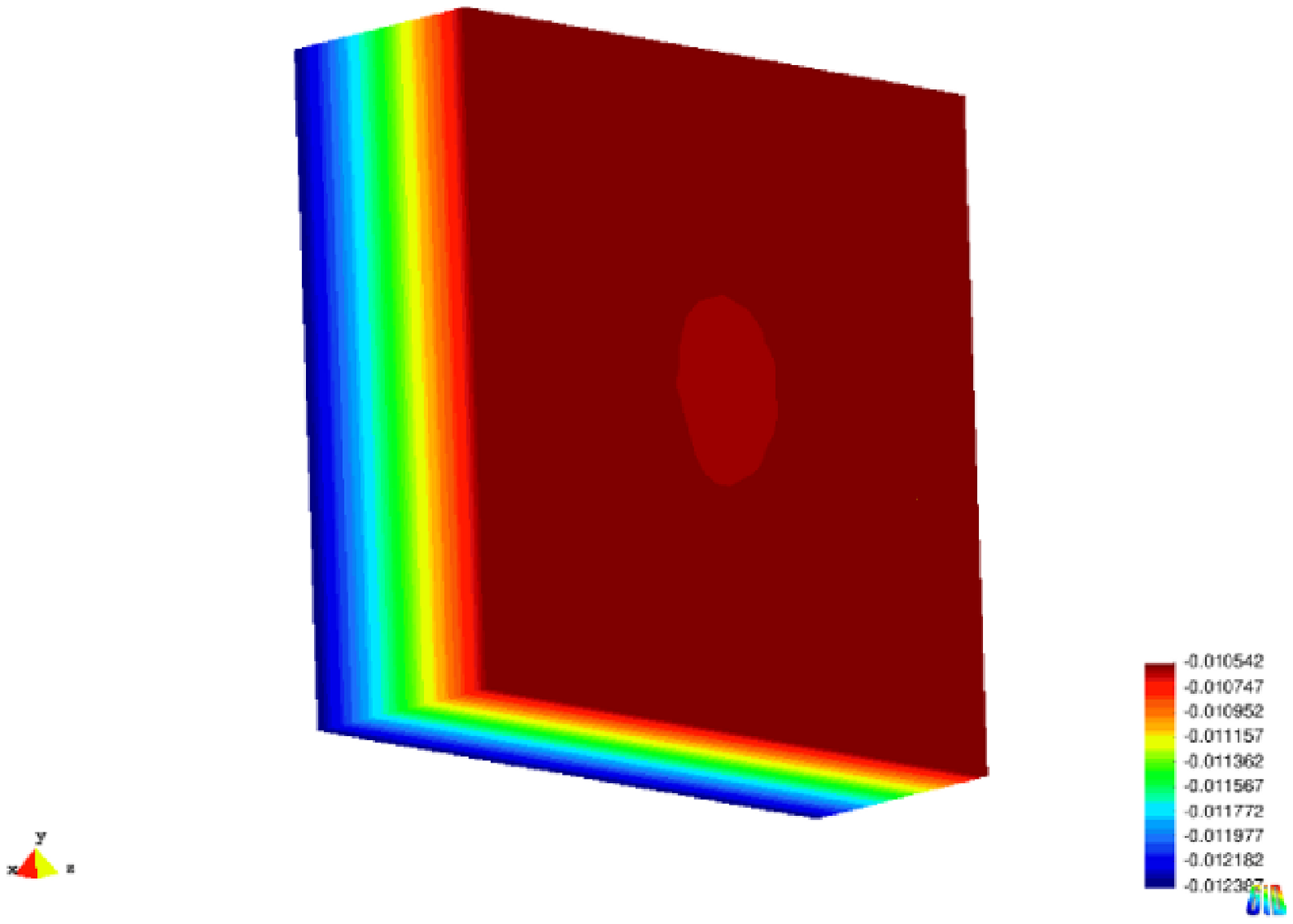}} &
 {\includegraphics[width=0.45\textwidth,clip=]{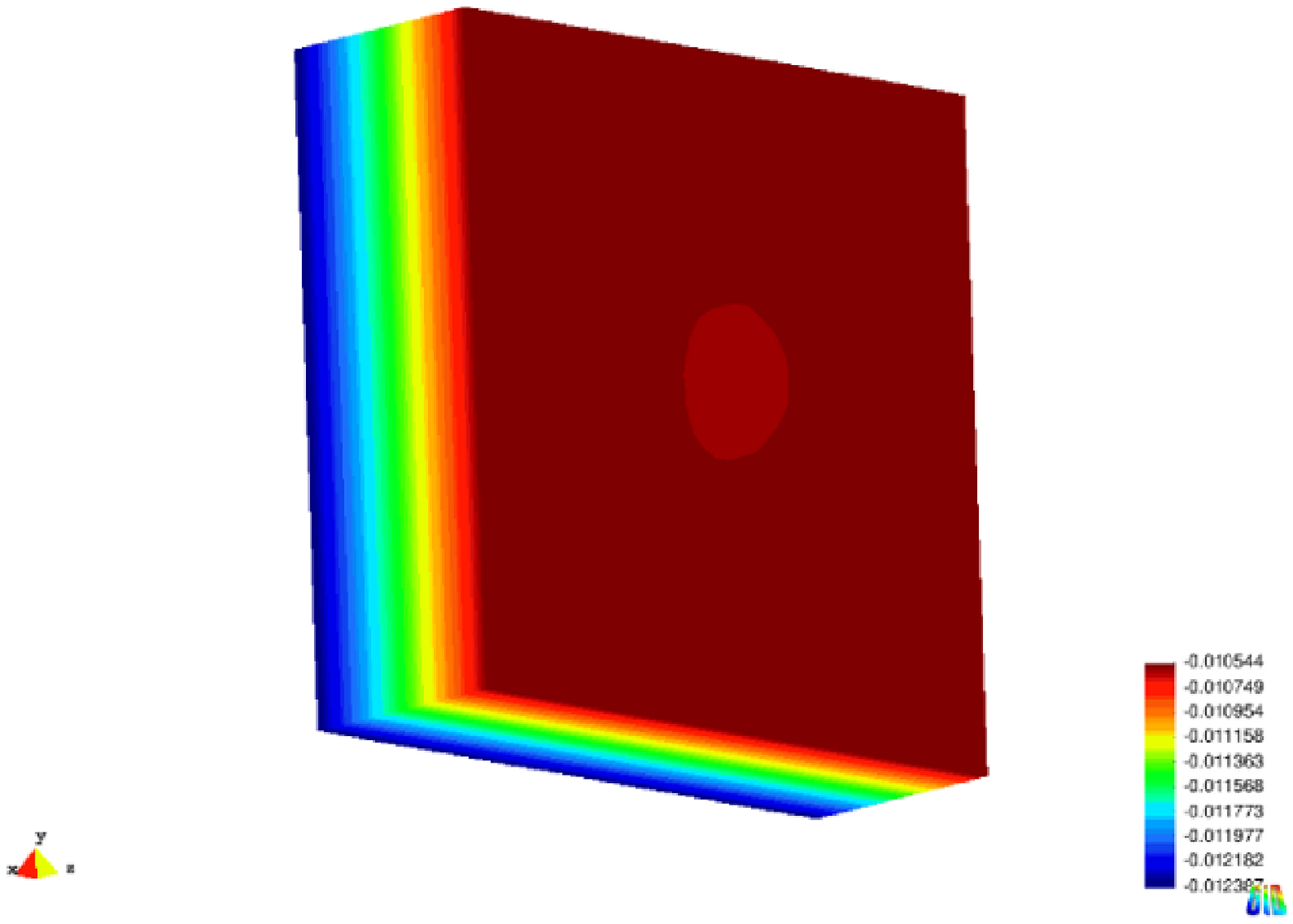}}\\
a)  object 1  & b)  object 2 \\
 {\includegraphics[width=0.45\textwidth,clip=]{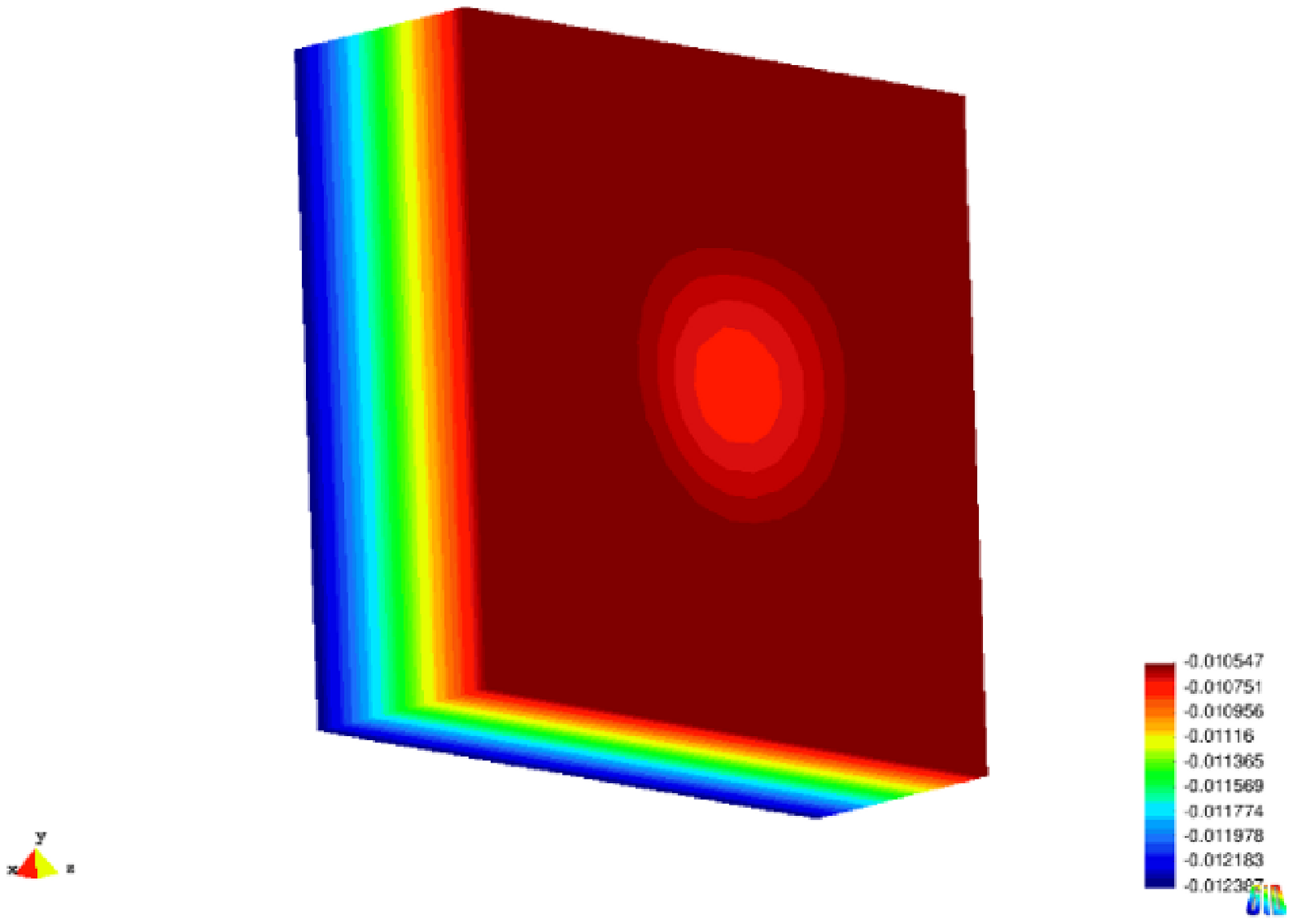}} &
{\includegraphics[width=0.45\textwidth,clip=]{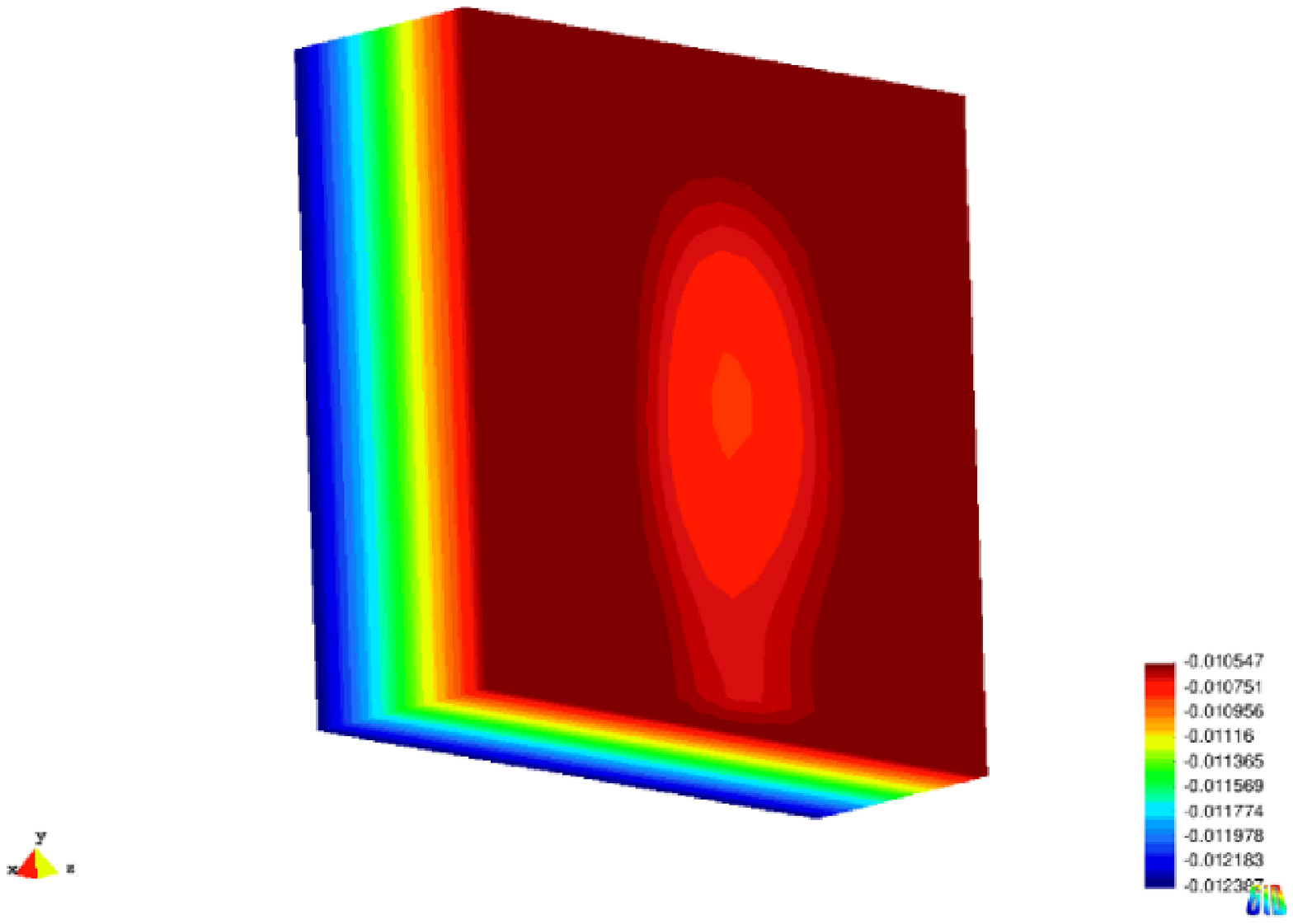}}
\\
c)  object 3  & d)  object 4 \\
{\includegraphics[width=0.45\textwidth,clip=]{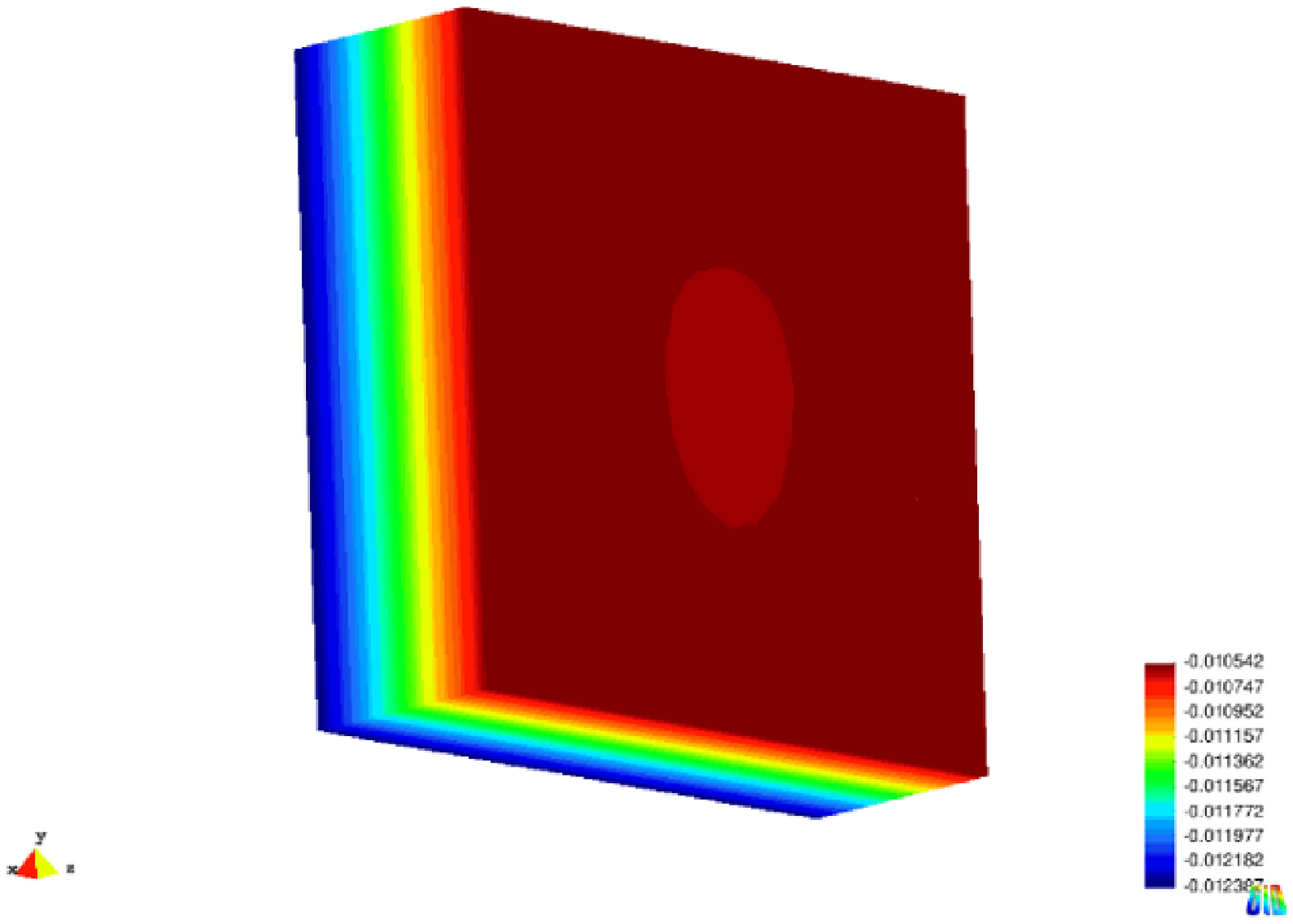}} &
 {\includegraphics[width=0.45\textwidth,clip=]{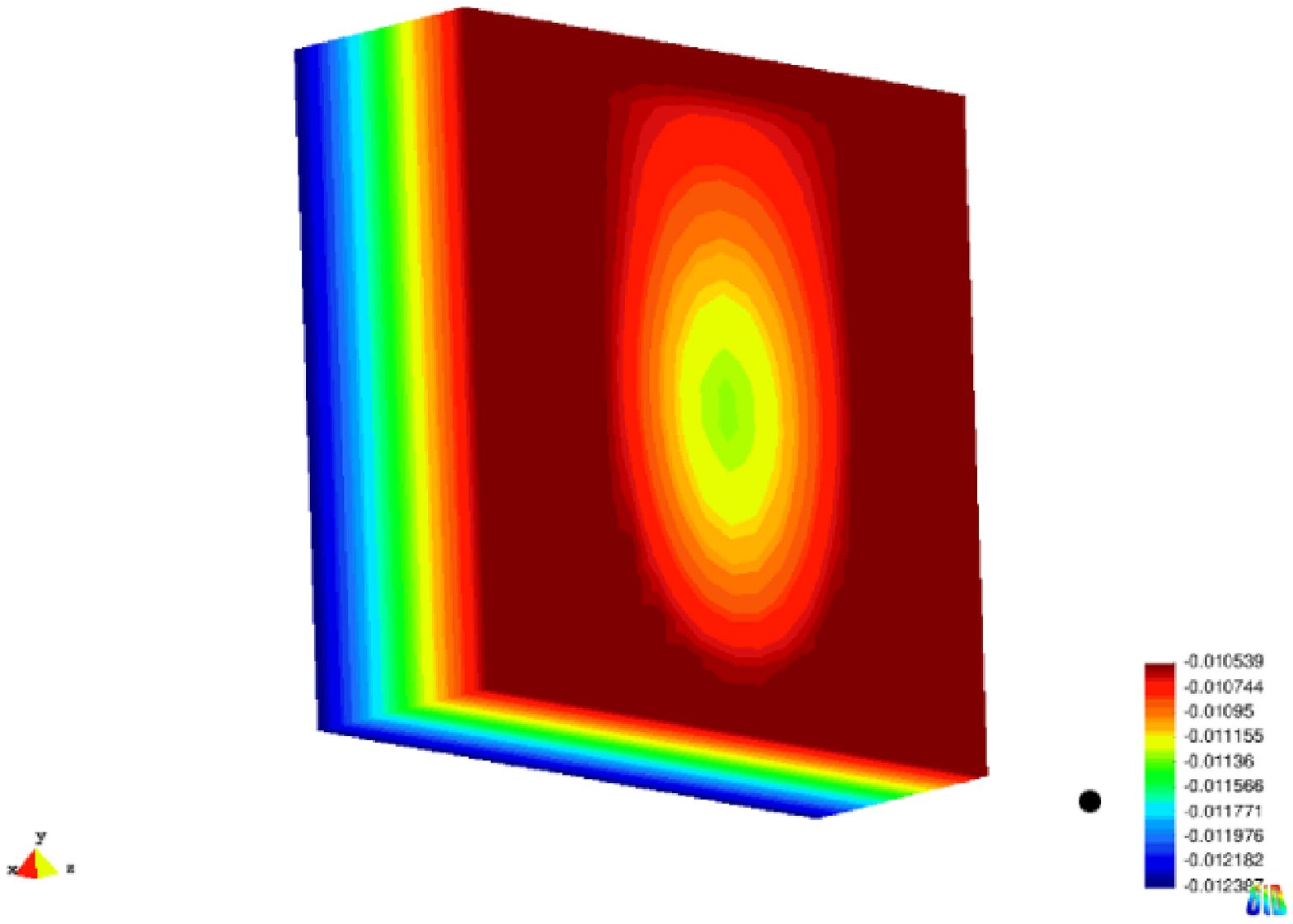}}\\
e)  object 5  & f)  object 6 \\
\end{tabular}
\end{center}
\caption{\protect\small \emph{ Behavior of functions
$\psi_n(x)$ at $\partial \Omega$ for some objects of Table
\ref{tab:table1}  at  pseudo-frequency $s=9.2$.}}
\label{fig:backscat_test4}
\end{figure}

\subsection{Reconstructions}

\label{sec:8.2}

We collected experimental data for 11 targets presented in  Table \ref{tab:table1}. Five targets were
dielectrics, five were metallic ones, and one was a metal covered by a
dielectric. We had total 7 blind cases: three dielectric, three metallic
targets and one target was the above mixture of the metal and a dielectric.
Three out of eleven targets were heterogeneous ones, all three were blind ones.
Heterogeneous targets model explosive devices in which explosive materials
are masked by dielectrics.

%We computationally simulated data for two non-blind targets: target
%number 1 (a piece of oak) and target number 3 (a metallic sphere) at
%the measurement plane $P_{m},$ using the same sizes as they actually
%have and assigning $\varepsilon _{r}=4.3$ for target number 1 and
%$\varepsilon _{r}=12$ for target number 3. Then we propagated these
%simulated data to the surface $\Gamma $ and applied the algorithm of
%section \ref{sec:6.1} to these propagated synthetic data. We have used
%mesh size step $h_{\mathbf{x}}=0.02$ in our computations with
%simulated data. Results (not shown) were about the same as ones
%presented below for targets number 1 and 3 for real data.

When proceeding with the algorithm of section \ref{sec:6.1}, we first
assign the Dirichlet boundary condition $\psi \left(
\mathbf{x},s\right) $ at $\partial \Omega $ for the function $q\left(
\mathbf{x},s\right) $ following (\ref{3.9}), (\ref{3.91}) and
(\ref{3.92}), in which case $\Gamma $ is as in (\ref{8.01}). Next, we
calculate functions $\psi_{n}\left( \mathbf{x}\right) $ as in
(\ref{4.21}).  Figure \ref{fig:backscat_test4} presents typical behavior of functions
$\psi_n(x)$ at $\partial \Omega$ for some objects of Table
\ref{tab:table1}. To have a better visualization, these figures are
zoomed to $0.4\times 0.4$ square from the $1\times 1$ square.

Table \ref{tab:table2} lists both computed $n^{\text{comp}}$ and
directly measured refractive indices $n$ of dielectric targets for
tests 1-3, see (\ref{4.300}) for $n^{\text{comp}}$. This table also
shows the measurement error in direct measurements of $n$. These
measurements were performed by the classical oscilloscope method
\cite{G}. Table \ref{tab:table3} lists computed appearing dielectric
constants $\varepsilon _{r}^{\text{comp}}$ of metallic targets. Recall
that $\varepsilon _{r}=n^{2}.$ We see from Table \ref{tab:table2} that
$\left( n^{\text{comp}}\right) ^{2} < 4.9$ for all dielectric
targets. This is going along well with the Step 4 of the stopping
criterion. On the other hand, in Table \ref{tab:table3} $\varepsilon
_{r}^{\text{comp}}>12$ for all metallic targets. Thus, our algorithm
    can confidently differentiate between dielectric and metallic
    targets.

\begin{table}
\begin{center}
\small
\begin{tabular}{|l|l|l|l|l|l|c|}
\hline
Target number & 1 & 2 & 5 & 8 & 10 & Average error \\ 
\hline  
blind/non-blind? & no & no & yes & yes & yes &  \\ 
\hline 
Measured $n$, error & 2.11, 19\% & 1.84, 18\% & 2.14, 28\% & 1.89, 30\% & 
2.1, 26\% & 24\% \\ 
\hline  
$n^{\text{comp}}$ of Test 1, error & 1.92, 10\% & 1.8, 2\% & 1.83, 17\% & 
1.86, 2\% & 1.92, 9\% & 8\% \\ 
\hline
$n^{\text{comp}}$ of Test 2, error & 2.07, 2\% & 2.01, 10\% & 2.21, 3\% & 
1.83, 3\% & 2.2, 5\% & 4.6\% \\ 
\hline 
$n^{\text{comp}}$ of Test 3, error & 2.017, 5\% & 2.013, 9\% & 2.03, 5\% & 
1.97, 4\% & 2.02, 4\% & 5\%  \\
\hline  
\end{tabular}
\end{center}
\caption{\emph{Computed $n^{\text{comp}}$ and directly measured
refractive indices of dielectric targets together with both measurement and
computational errors as well as the average error. Note that the average computing errors are at least three times less than the average error of direct measurements. }}
\label{tab:table2}
\end{table}

One can derive several important observations from Table \ref{tab:table2}.
First, in all three tests and for all targets the computational error is
significantly less than the error of direct measurements. Thus, the average
computational error is significantly less than the average measurement error
in all three tests. Second, computed refractive indices are within trust
intervals in all cases. The accuracy of all three tests is about the same.

Table \ref{tab:table3} provides information about computed appearing
dielectric constants $\varepsilon _{r}^{\text{comp}}$ of metallic
targets, see (\ref{2.51}) and (\ref{4.300}). Note that in Test 3 first
four numbers $\varepsilon _{r}^{\text{comp}}=15$. This coincides with
the upper bound in (\ref{8.20}). On the other hand, $\varepsilon
_{r}^{\text{comp}}=14 < 15$ for the target number 9.\ This is probably
because target number 9 is a mixture of a metal and a dielectric. An
important observation, which can be derived from Table
\ref{tab:table2}, is that our algorithm confidently computes large
inclusion/background contrasts exceeding 10:1. It is well known that
optimization methods of conventional least squares residual
functionals usually cannot image large contrasts.

\begin{table}
\begin{center}
\begin{tabular}{|l|l|l|l|l|l|l|l|}
\hline
Target number & 3 & 4 & 6 & 7 & 9 & 11\\ 
\hline
blind/non-blind (yes/no) & no & no & yes & yes & yes & yes \\ 
\hline
$\varepsilon _{r}^{\text{comp}}$ of Test 1 & 14.4 & 15.0 & 15 & 13.6 & 13.6
& 13.1\\ 
\hline
$\varepsilon _{r}^{\text{comp}}$ of Test 2 & 15 & 15 & 15 & 14.1 & 14.1
& 15 \\ 
\hline
$\varepsilon _{r}^{\text{comp}}$ of Test 3 & 15 & 15 & 15 & 15 & 14 & 14.06 \\
\hline
\end{tabular}
\end{center}
\caption{\emph{Computed appearing dielectric constants }$\varepsilon
  _{r}^{\text{comp}}$\emph{\ of metallic targets number 3,4,6,7,11 as
    well as of the target number 9 which is a metal covered by a
    dielectric.}}
\label{tab:table3}
\end{table}

\begin{figure}[tbp]
\begin{center}
\begin{tabular}{cc}
{\includegraphics[scale=0.18, angle=-90, clip=]{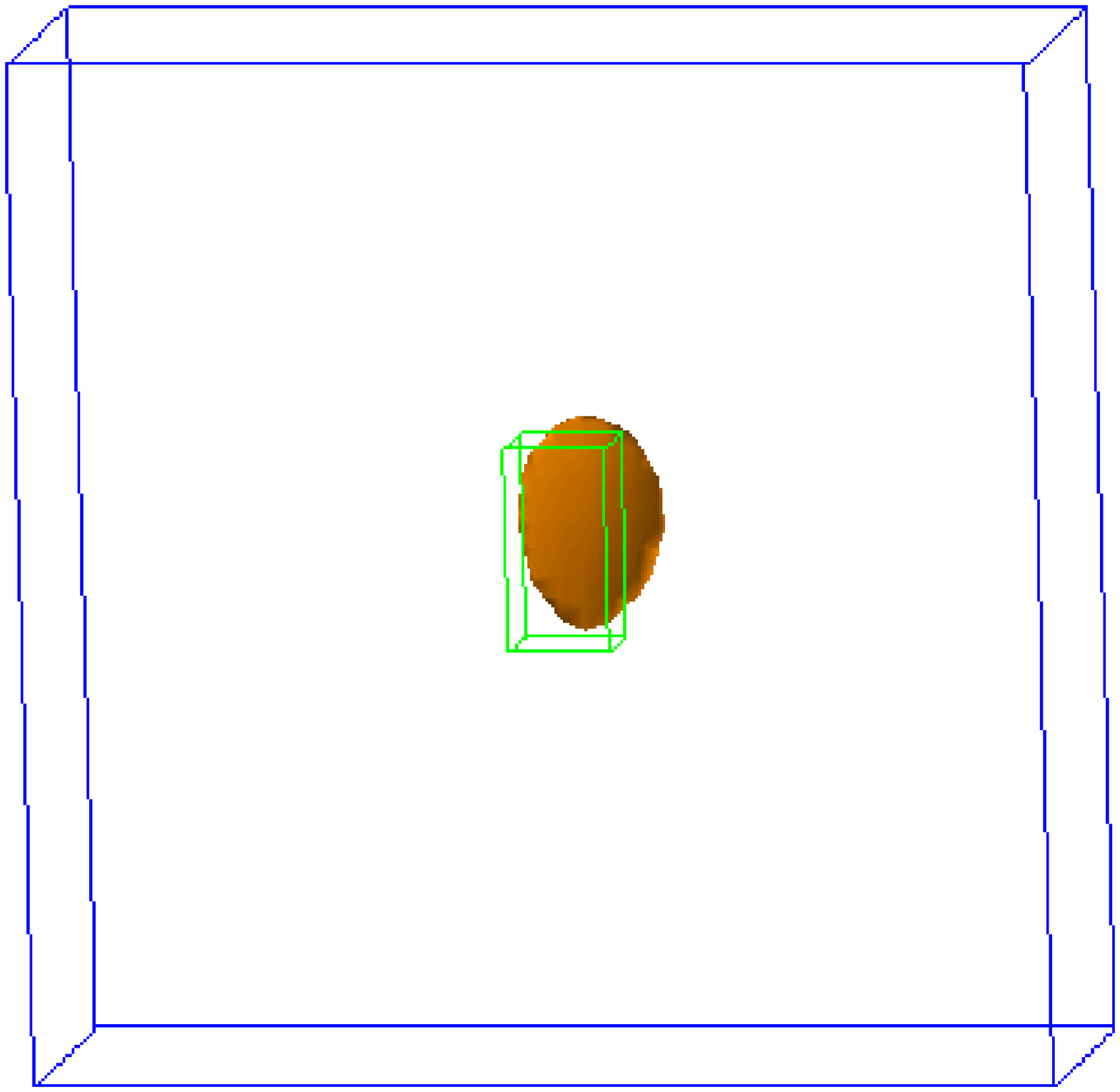}} &
{\includegraphics[scale=0.18, angle=-90, clip=]{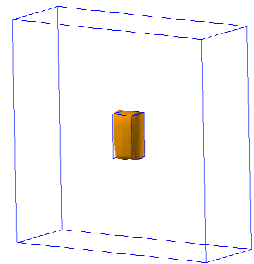}} \\
a) Test 1 (dielectric), object 1, first stage &  b)   Test 1 (dielectric), object 1, second  stage\\
{\includegraphics[scale=0.18, angle=-90, clip=]{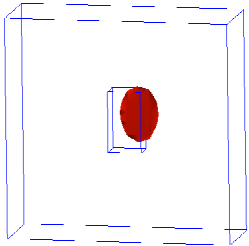}} &
{\includegraphics[scale=0.18, angle=-90, clip=]{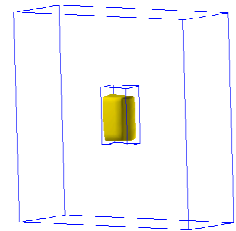}} \\
c) Test 1 (dielectric), object 5, first stage &  d)   Test 1 (dielectric), object 5, second  stage\\
\end{tabular}
\end{center}
\caption{{\protect\small \emph{Computed images of targets numbers 1,5 of Table \ref{tab:table1}. Thin
      lines indicate correct shapes. To have better visualization we
      have zoomed images of Tests 1,2 from the domain $\Omega$ defined
      by (\ref{8.0}) to the domain (\ref{8.1}). }}}
\label{fig:fig10}
\end{figure}

All targets, except of targets number 8, 9, 10, were homogeneous ones
comprised from a single substance only. However, targets number 8-10
were inhomogeneous ones,  see Table \ref{tab:table1} for description
of all targets. The target number 8 was a wooden doll which was empty
inside. In the case of target number 9, a piece of a metal was
inserted inside that doll. Thus, only the metal was imaged, because
its reflection is much stronger than the wood. In the case of target
number 10, sand was partly inserted inside that doll.

\begin{figure}[tbp]
\begin{center}
\begin{tabular}{ccc}
{\includegraphics[scale=0.21, angle=-90, clip=]{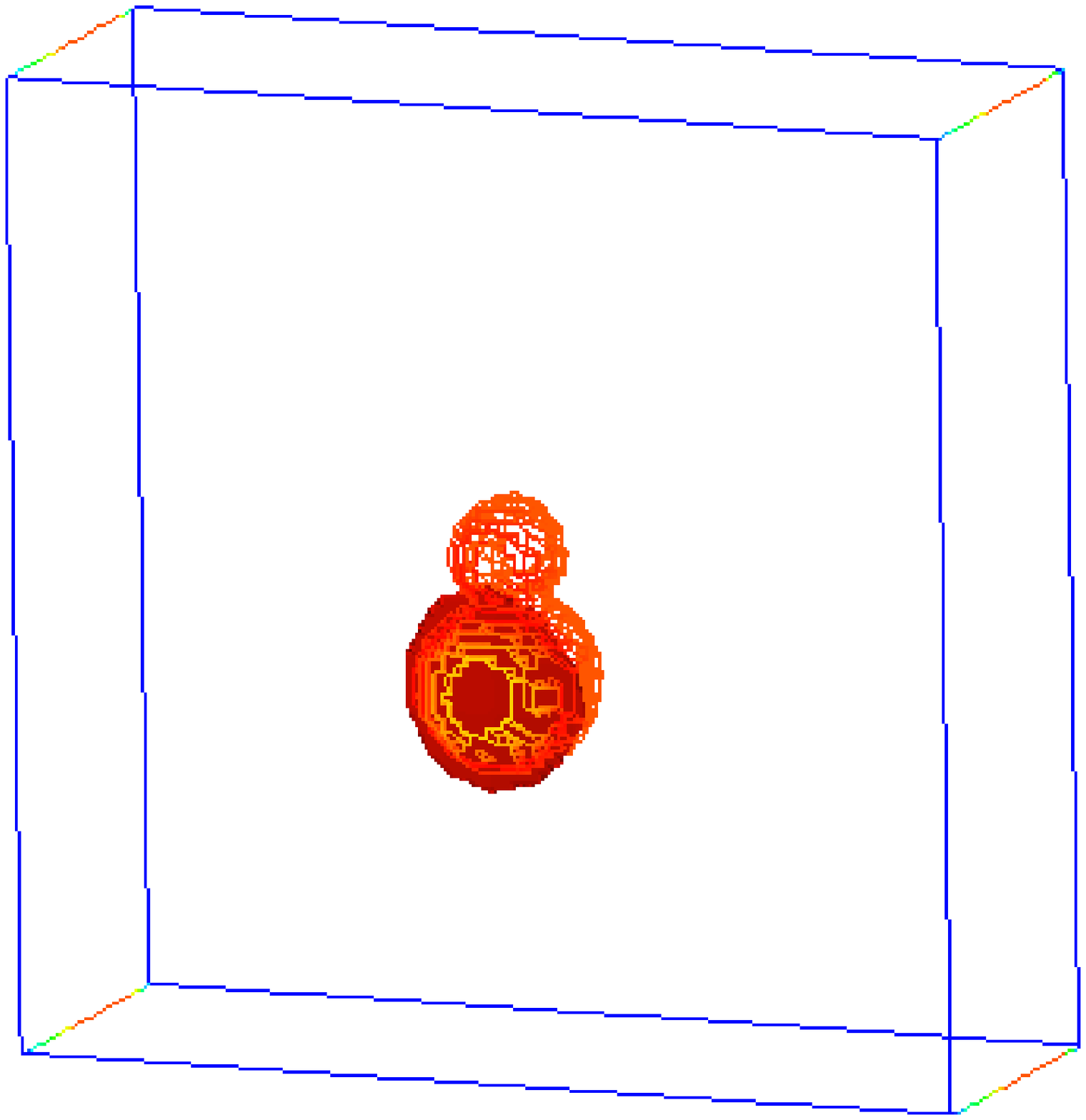}} &
{\includegraphics[scale=0.19, angle=-90, clip=]{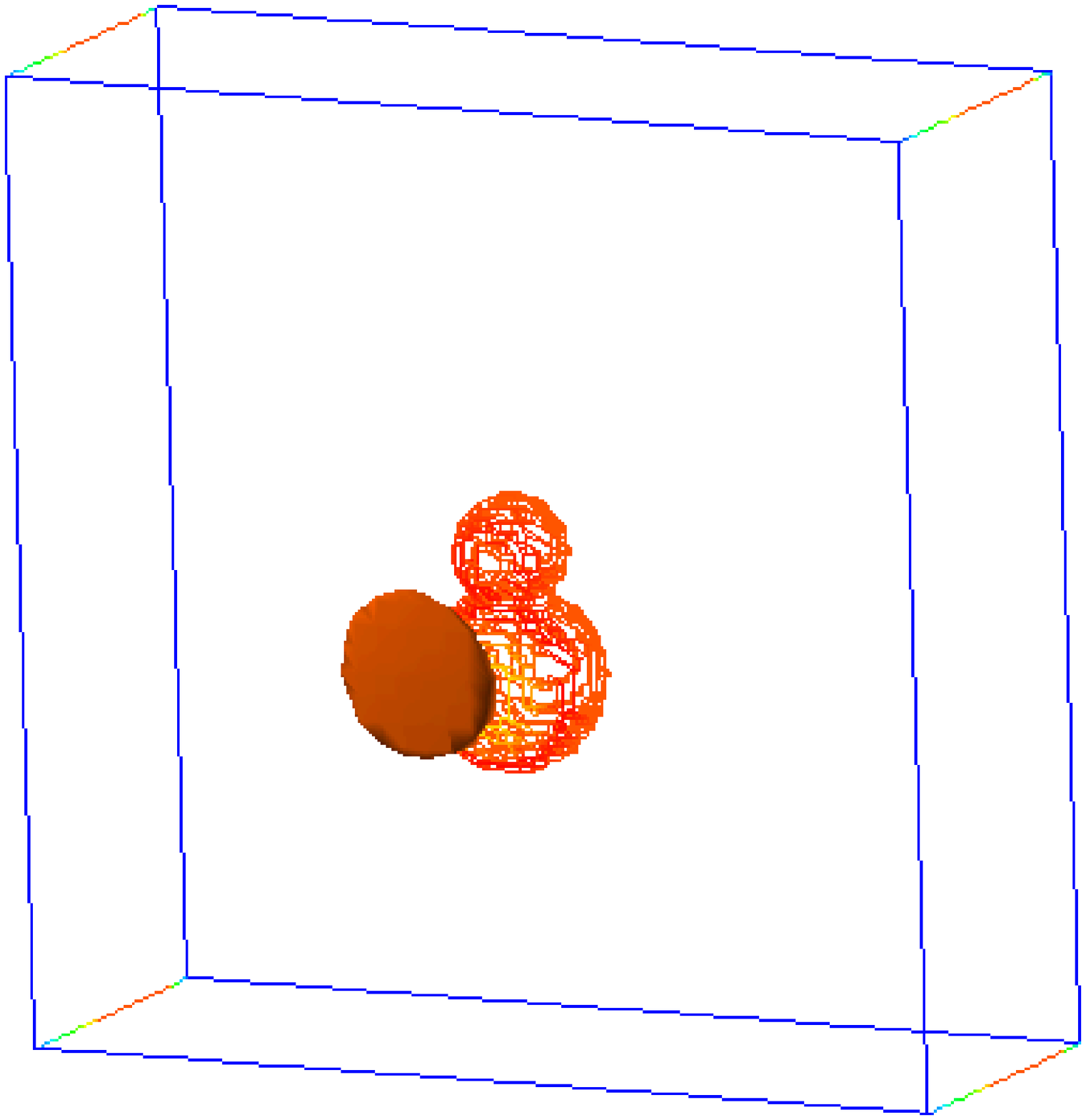}}  &
 {\includegraphics[scale=0.19, angle=-90, clip=]{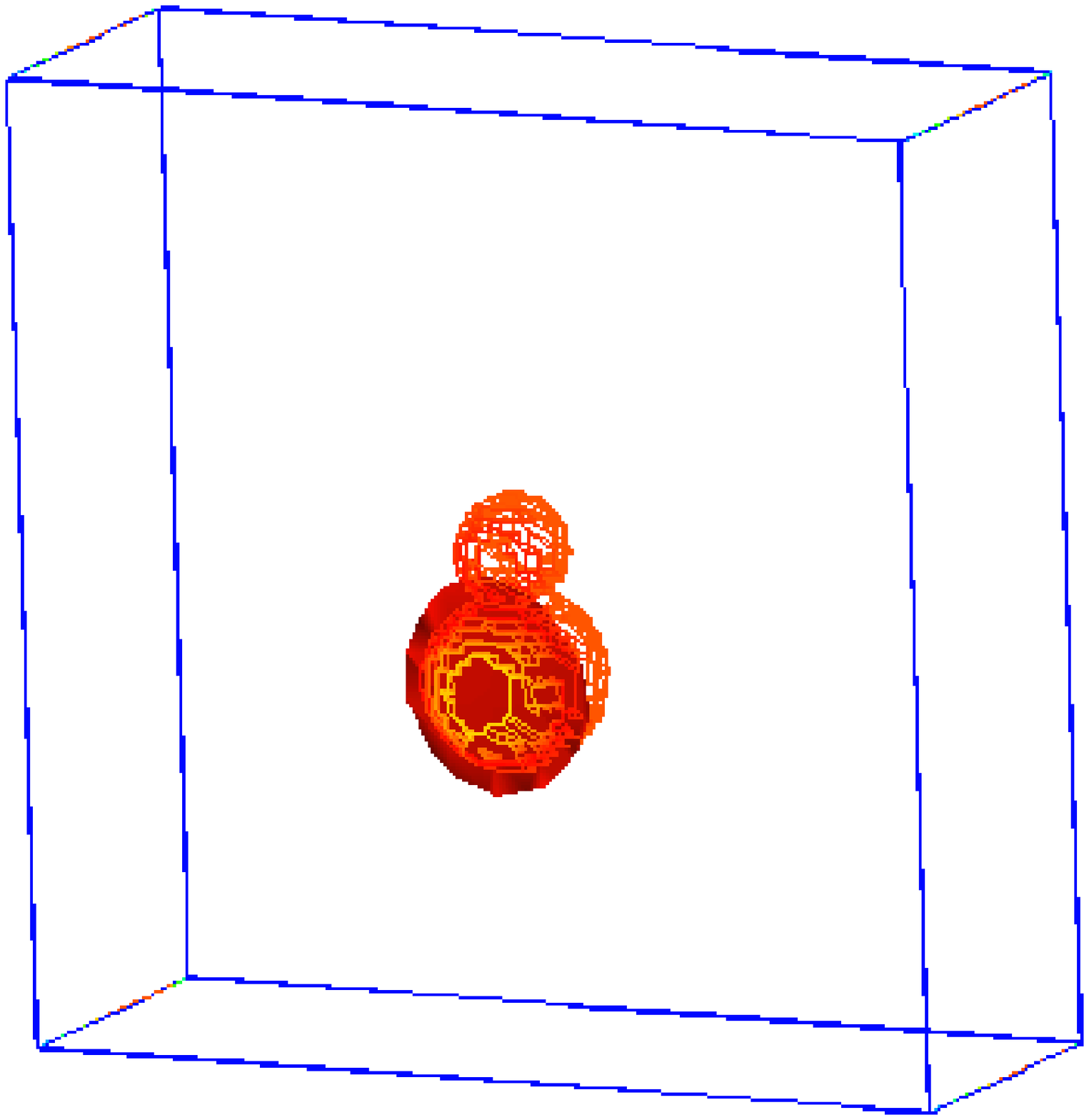}}  \\
a) Test 1, object 8 &  b)   Test 2, object 8  &  c)   Test 3, object 8\\
{\includegraphics[scale=0.21, angle=-90, clip=]{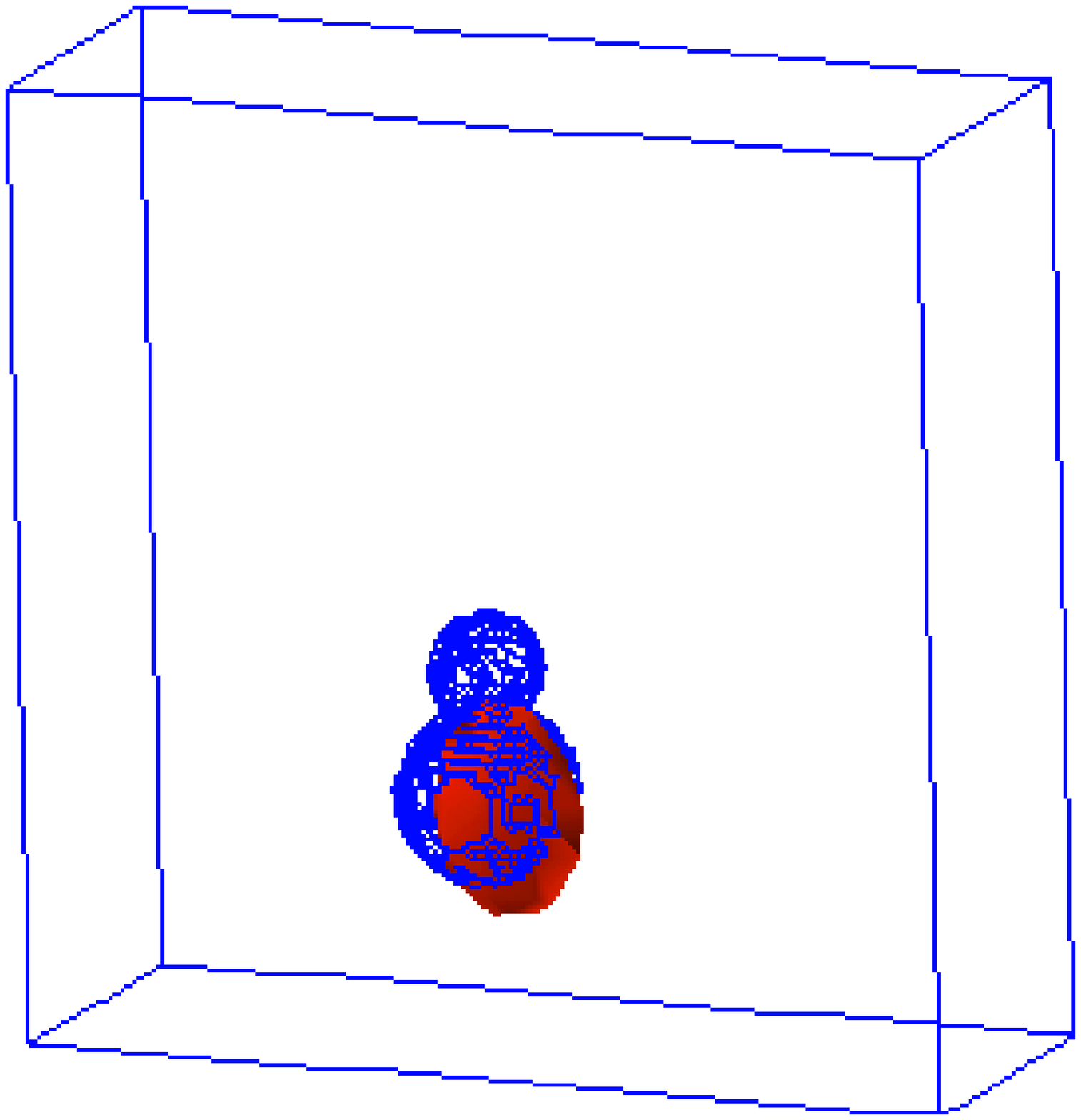}} &
{\includegraphics[scale=0.21, angle=-90, clip=]{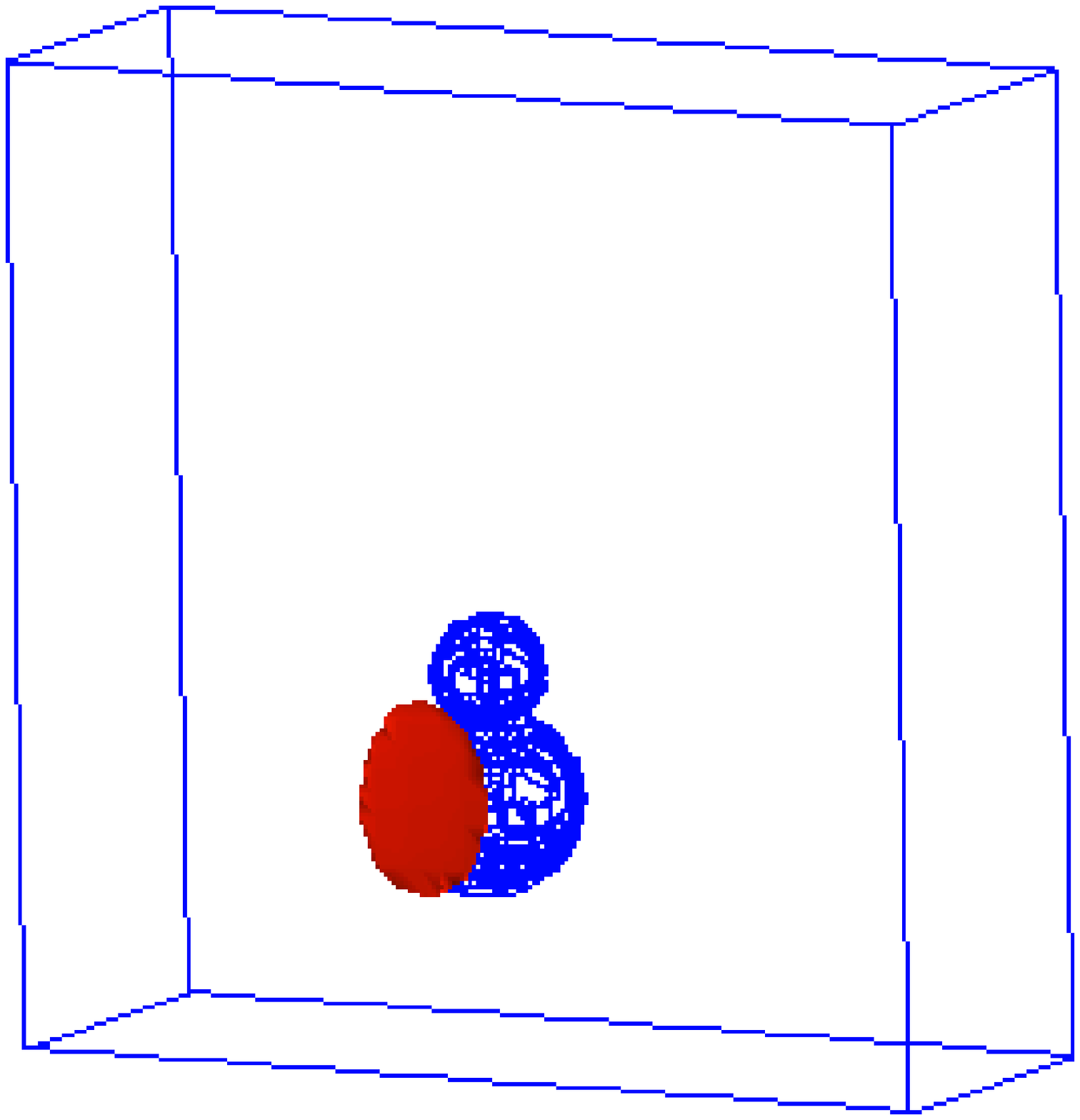}}  &
 {\includegraphics[scale=0.21, angle=-90, clip=]{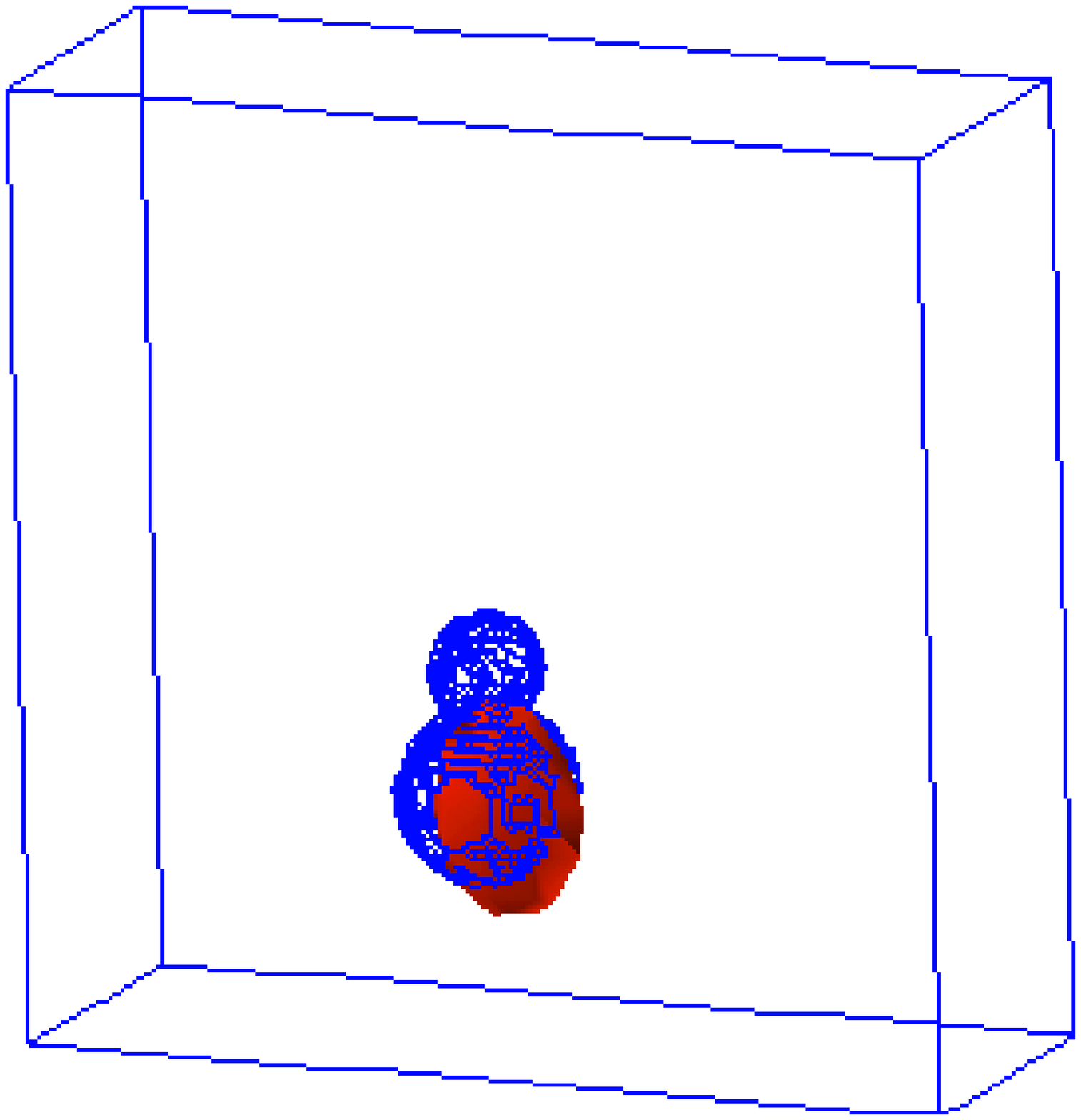}}  \\
d) Test 1, object 9 &  e)   Test 2, object 9  &  f)   Test 3, object 9\\
{\includegraphics[scale=0.21, angle=-90, clip=]{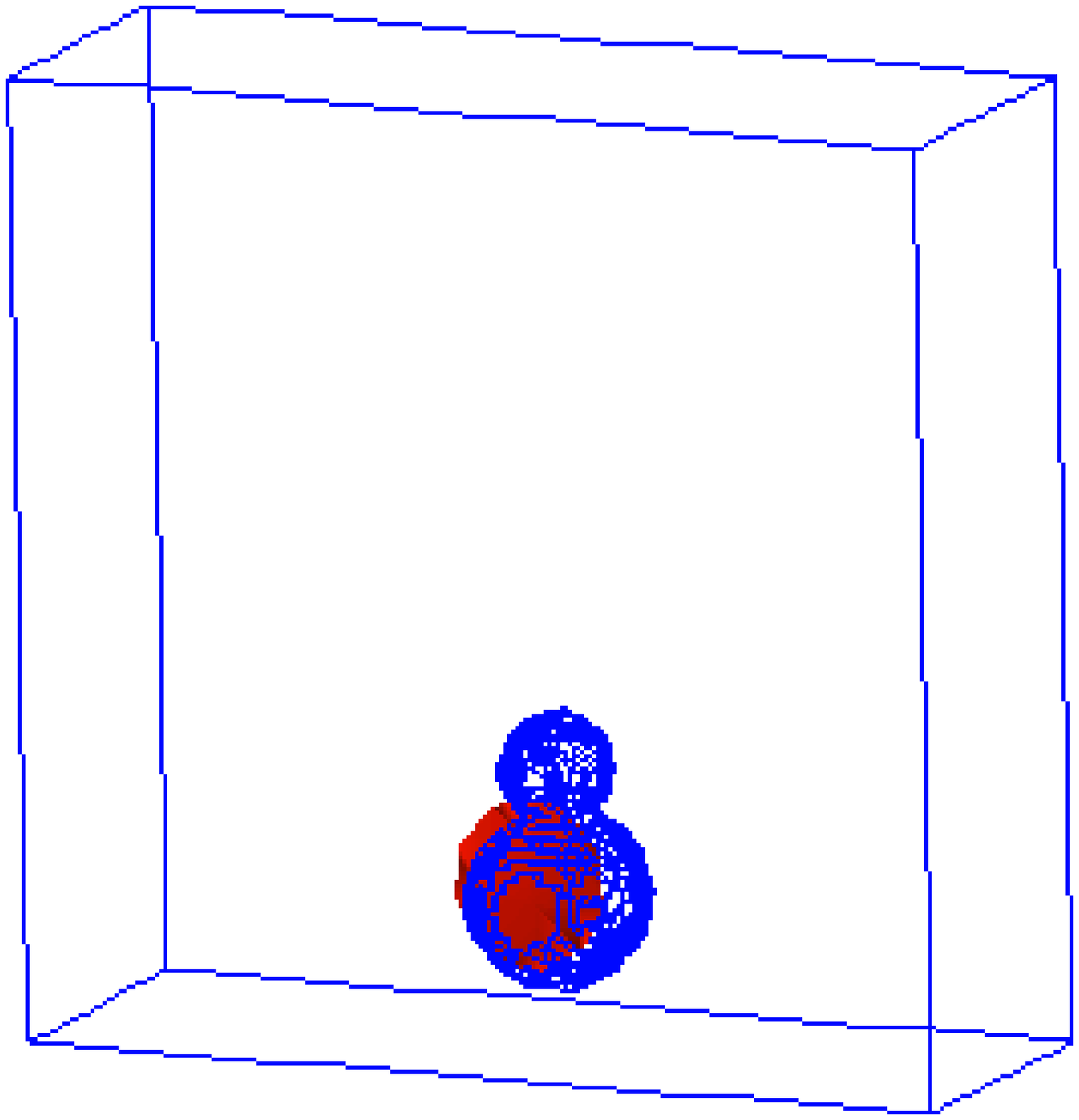}} &
{\includegraphics[scale=0.21, angle=-90, clip=]{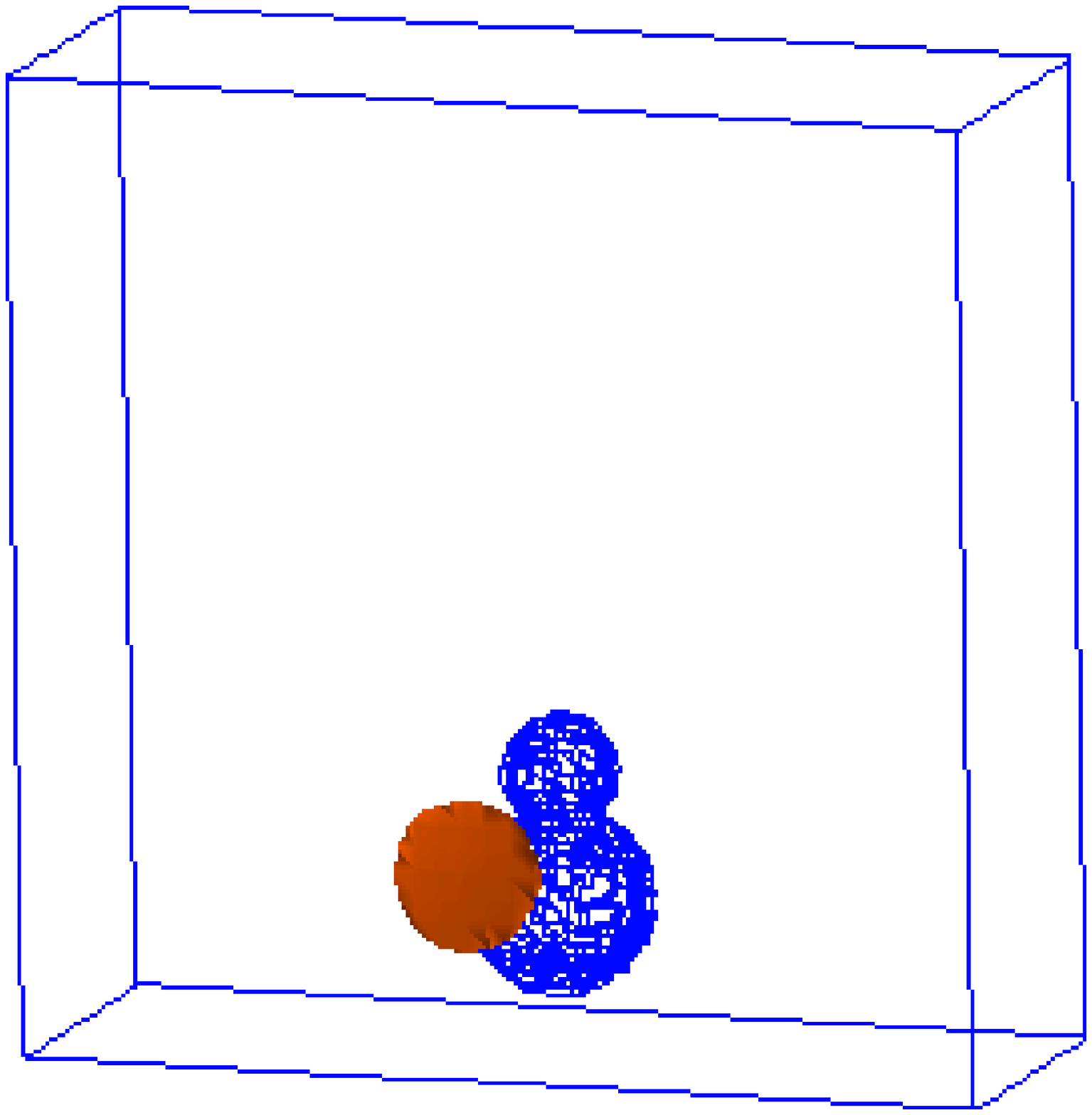}}  &
 {\includegraphics[scale=0.21, angle=-90, clip=]{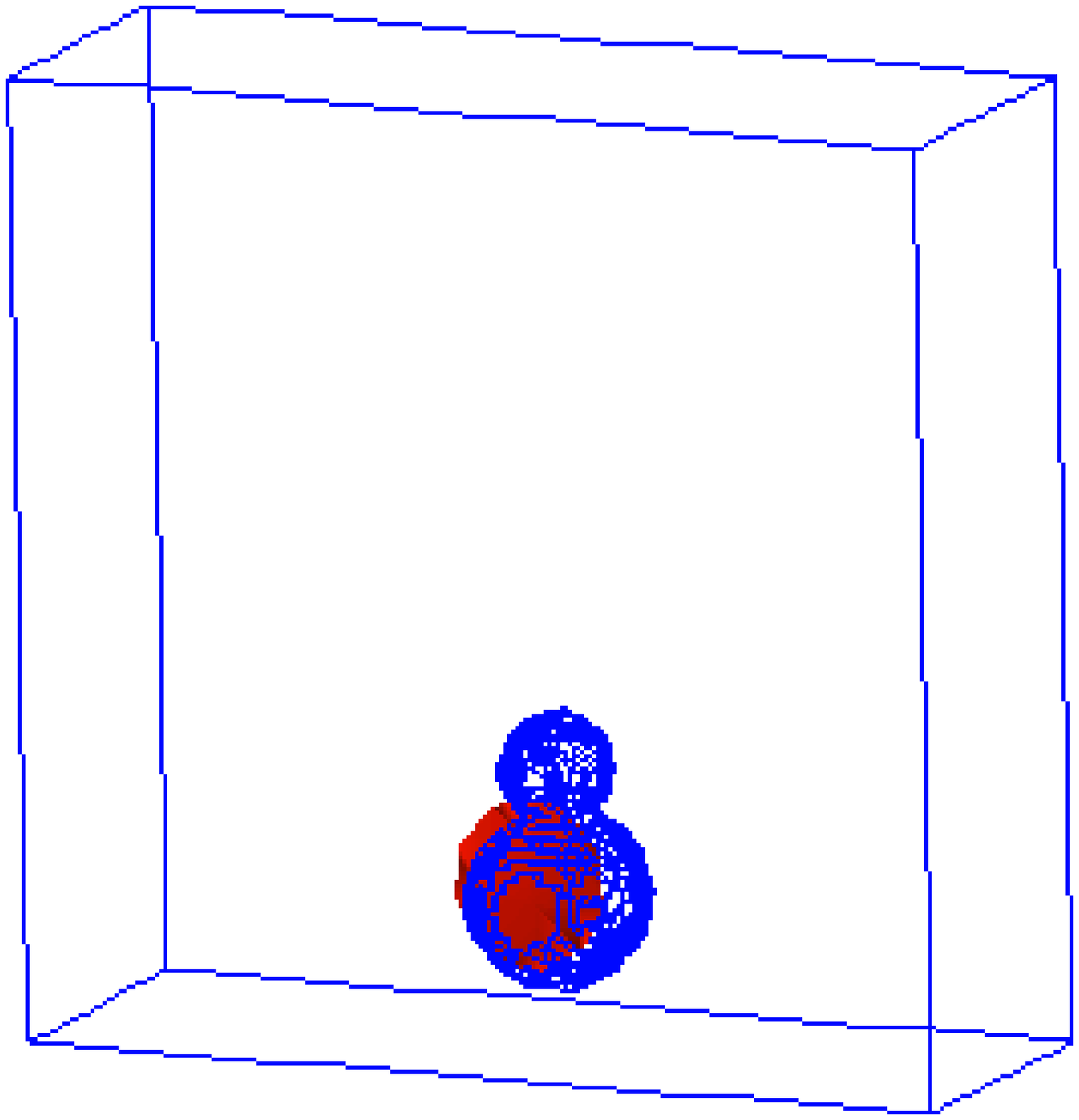}}  \\
g) Test 1, object 10 &  h)   Test 2, object 10  &  i)   Test 3, object 10\\
\end{tabular}
\end{center}
\caption{{\protect\small \emph{Computed images of targets numbers
      8,9,10 (see Table \ref{tab:table1}). Thin lines indicate correct
      shapes. To have better visualization we have zoomed images of
      Tests 1,2 from the domain $\Omega$ defined by (\ref{8.0}) to the
      domain (\ref{8.1}). }}}
\label{fig:fig11}
\end{figure}

\begin{figure}[tbp]
\begin{center}
\begin{tabular}{ccc}
{\includegraphics[scale=0.21, angle=-90, clip=]{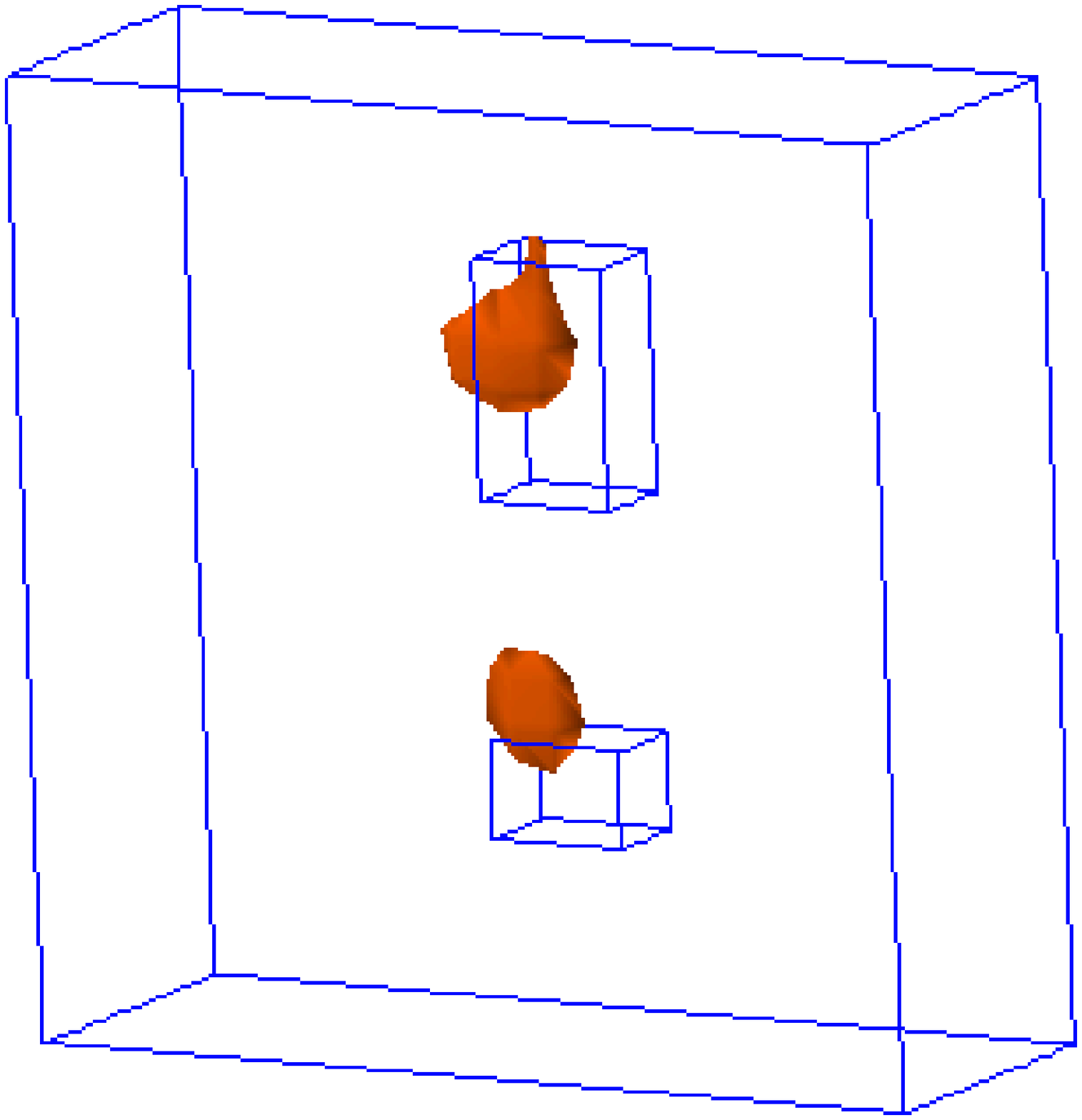}} &
{\includegraphics[scale=0.19, angle=-90, clip=]{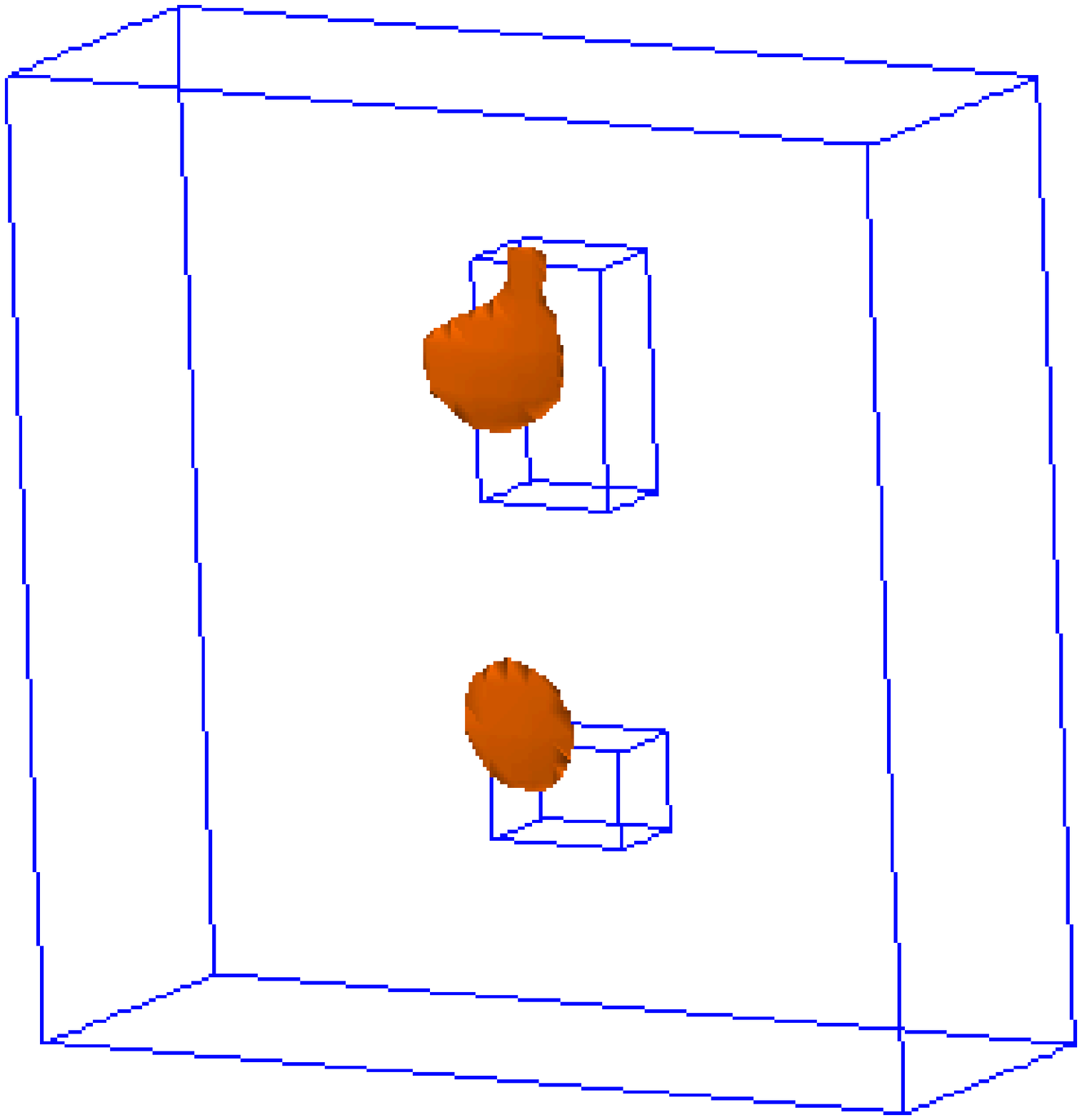}}  &
 {\includegraphics[scale=0.19, angle=-90, clip=]{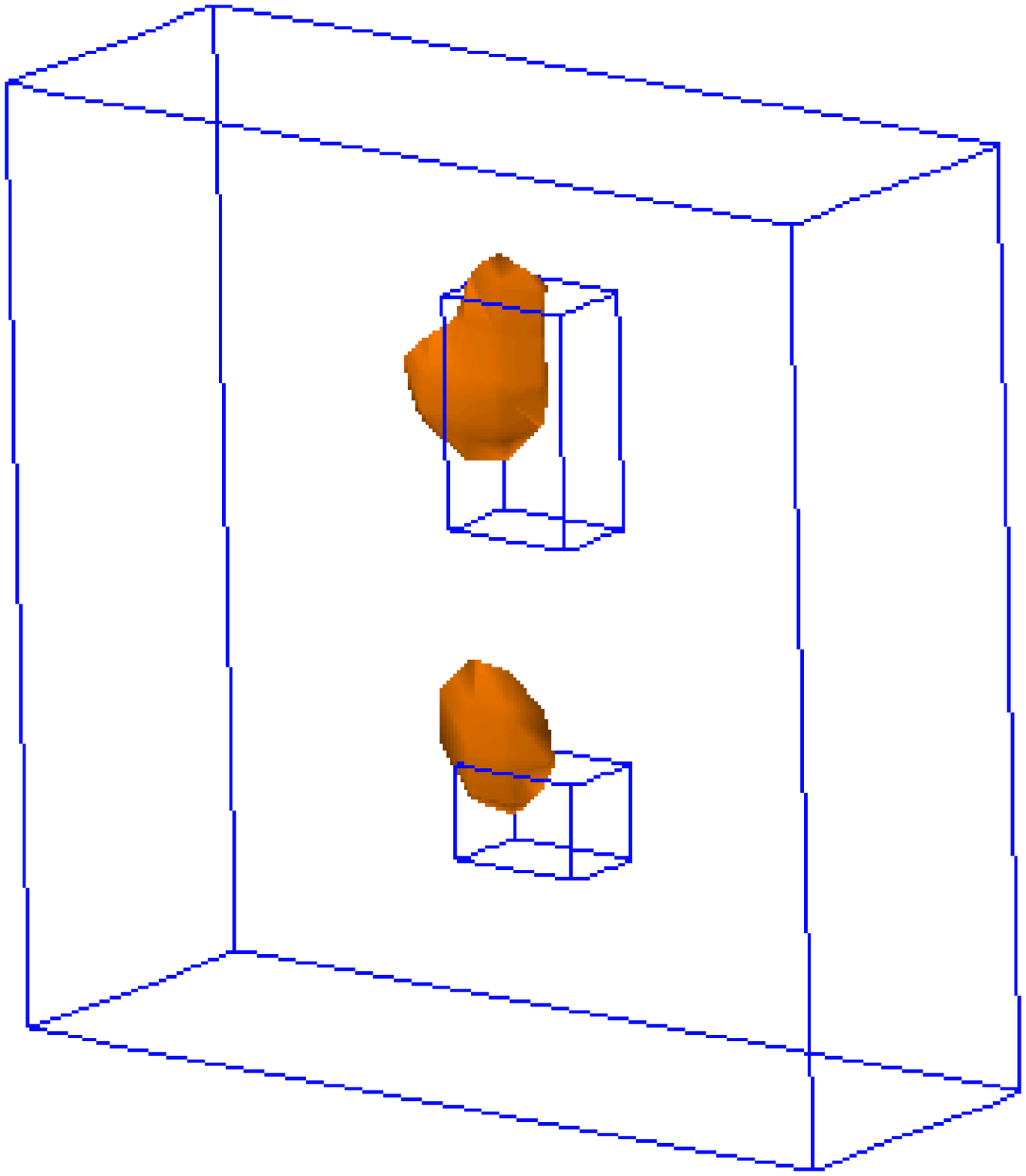}}  \\
a) Test 1, object 11 &  b)   Test 2, object 11  &  c)   Test 3, object 11 \\
\end{tabular}
\end{center}
\caption{{\protect\small \emph{Computed images of target number
      11 of Table \ref{tab:table1}. Thin lines indicate correct shapes
      of two inclusions to be reconstructed. To have better
      visualization we have zoomed images of Tests 1,2 from the domain
      $\Omega$ defined by (\ref{8.0}) to the domain (\ref{8.1}). }}}
\label{fig:fig12}
\end{figure}

Figures \ref{fig:fig10} display 3-d images of some targets for Test 1
after the first and the second (postprocessing) stages described in
section \ref{sec:7.2}.  Figures \ref{fig:fig11}, \ref{fig:fig12} display 3-d images of
targets 8,9,10 and 11 for all three tests.

Note that it is hard to estimate well the size of a target in the
$z-$direction.  Nevertheless, one can observe that rather good shapes
and sizes of targets are computed in the case of prisms and cylinders,
see Figure \ref{fig:fig10}.  As to the doll, neither of tests images
shapes of targets 8-10 accurately. Still, the location of the doll
\ as well as its sizes in $x,y$ directions are well estimated, see Figures \ref{fig:fig11}.

\section{Summary}

\label{sec:9}

We collected experimental backscattering time resolved data of electrical
wave propagation and have applied the approximately globally convergent
numerical method of \cite{BK} to these data. Results for four non-blind and
seven blind cases show a good accuracy of reconstruction of refractive indices
of dielectric targets and appearing dielectric constants of metallic
targets. In the case of dielectrics, the average reconstruction error is at
least three times less than the error of direct measurements. We confidently
differentiate between metallic and dielectric targets. In particular, we
have accurately computed maximal values of refractive indices/dielectric
constants of three blind heterogeneous targets. These targets represent
simplified models of improvised explosive devices IEDs, which are
heterogeneous ones.

Locations of targets and their sizes in $x,y$ directions are
accurately reconstructed. The most difficult cases of sizes in the
$z-$direction (depth) are well reconstructed in some cases. In
addition, shapes of some targets are well reconstructed in some
cases. We believe that a follow up application of the locally
convergent adaptivity technique might improve reconstructions of
shapes of targets. The adaptivity takes the solution obtained by the
approximately globally convergent method as the starting point for the
minimization of the Tikhonov functional on a sequence of adaptively
refined meshes. A significant refinement via the adaptivity was
demonstrated in section 5.9 of \cite{BK} for the case of transmitted
experimental data, see Figures 5.13 and 5.16 in \cite{BK}.

\begin{center}
\textbf{Acknowledgments}
\end{center}

This research was supported by US Army Research Laboratory and US Army
Research Office grants W911NF-11-1-0325 and W911NF-11-1-0399, the Swedish
Research Council, the Swedish Foundation for Strategic Research (SSF) through the
Gothenburg Mathematical Modelling Centre (GMMC) and by the Swedish
Institute, Visby Program. The authors are grateful to Mr. Steven Kitchin for
his excellent work on data collection.


\begin{thebibliography}{99}


\bibitem{BakKok} A.B.\ Bakushinsky and M.Yu. Kokurin, \emph{Iterative
Methods for Approximate Solutions of Inverse Problems}, Springer, New York,
2004.

\bibitem{BSA} L. Beilina, K. Samuelsson and K. Ahlander, Efficiency of a
hybrid method for the wave equation. In \emph{\ International Conference on
Finite Element Methods}, Gakuto International Series Mathematical Sciences
and Applications, Gakkotosho CO., LTD, 2001.

\bibitem{BK} L. Beilina and M.V. Klibanov, \emph{Approximate Global
Convergence and Adaptivity for Coefficient Inverse Problems}, Springer, New
York, 2012.

\bibitem{BKJIIP12} L. Beilina and M.V. Klibanov, A new approximate
mathematical model for global convergence for a coefficient inverse problem
with backscattering data, \emph{J. Inverse and Ill-Posed Problems}, 20,
513-565, 2012.

\bibitem{BM} L.\ Beilina, Energy estimates and numerical verification of the
stabilized domain decomposition finite element/finite difference approach
for the Maxwell's system in time domain, \emph{Central European Journal of
Mathematics}, 11, 702-733, 2013.

\bibitem{BukhK} A.L. Bukhgeim and M.V. Klibanov, Uniqueness in the large of
a class of multidimensional inverse problems, \emph{Soviet Math. Doklady},
17, 244-247, 1981.

\bibitem{Chav} G. Chavent, \emph{Nonlinear Least Squares for Inverse
Problems.\ Theoretical Foundations and Step-By-Step Guide for Applications},
Springer,\ New York, 2009. 

\bibitem{EHN} H.W. Engl, M.\ Hanke and A.\ Neubauer, Regularization of
Inverse Problems, Kluwer Academic Publishers, Boston, 2000.

\bibitem{EM} B. Engquist and A. Majda, Absorbing boundary conditions for the
numerical simulation of waves, \emph{\ Math. Comp.,} 31, 629-651, 1977.

\bibitem{G} H.H.\ Gerrish, W.E.Jr. Dugger and R.M. Robert, \emph{Electricity
and Electronics},\ Goodheart-Wilcox Co. Inc., Merseyside, UK, 2004.

\bibitem{Isakov} V.\ Isakov, Inverse obstacle problems, \emph{Inverse
Problems}, 25, 123002, 2009.

\bibitem{Johnson} C. Johnson,\ \emph{Numerical Solution of Partial
Differential Equations by the Finite Element Method}, Cambridge University
Press, Cambridge, 1987.

\bibitem{KBKSNF} A.V. Kuzhuget, L. Beilina, M.V. Klibanov, A. Sullivan,\ L.
Nguyen and M.A. Fiddy, Blind experimental data collected in the field and an
approximately globally convergent inverse algorithm, \emph{Inverse Problems}%
, 28, 095007, 2012.

\bibitem{LU} O.A.\ Ladyzhenskaya and N.N.\ Uralceva, \emph{Linear and
Quasilinear Elliptic Equations}, Academic Press,\ New York, 1969.

\bibitem{NBKF} N.~T. Th\`anh, L.~Beilina, M.~V. Klibanov, and
  M.~A. Fiddy.  Reconstruction of the refractive index from
  experimental backscattering data using a globally convergent inverse
  method, Preprint, arXiv:1306.3150 [math-ph], 2013.



\bibitem{Soumekh} M. Soumekh, \emph{Syntetic Aperture Radar Signal Processing%
}, Willey\&Son, New York, 1999.

\bibitem{Vog} C.R. Vogel, \emph{Computational Methods for Inverse Problems},
SIAM Publications, Philadelphia, 2002.

\bibitem{waves} WavES, the software package, http://www.waves24.com
\end{thebibliography}
\end{document}